\documentclass[a4paper,fleqn,usenatbib,useAMS,usedcolumns]{mnras}
\usepackage{newtxtext,newtxmath}

\usepackage[T1]{fontenc}
\usepackage{ae,aecompl}
\usepackage{comment}
\usepackage{threeparttable}
\usepackage{booktabs, caption, makecell}

\usepackage{graphicx}	
\usepackage{amsmath}	
\usepackage{gensymb}
\usepackage{siunitx}

\newcommand{\hi}{\mbox{H\,{\sc i}}}
\newcommand{\mgii}{\mbox{Mg\,{\sc ii}}}
\newcommand{\mgi}{\mbox{Mg\,{\sc i}}}

\newcommand{\siii}{\mbox{Si\,{\sc ii}}}

\newcommand{\civ}{\mbox{C\,{\sc iv}}}

\newcommand{\nv}{\mbox{N\,{\sc v}}}
\newcommand{\alii}{\mbox{Al\,{\sc ii}}}
\newcommand{\siv}{\mbox{Si\,{\sc iv}}}

\newcommand{\Anand}[1]{{\color{red}[Anand: #1]}}
\newcommand{\PPJ}[1]{{\color{green}[PPJ: #1]}}

\newcommand{\RN}[1]{%
  \textup{\uppercase\expandafter{\romannumeral#1}}%
}

\def\h2{$\rm H_2$}
\def\Nh2{$N$(H${_2}$)}

\def\lya{\ensuremath{{\rm Ly}\alpha}}

\def\kms{km\,s$^{-1}$}

\def\zabs{$z_{\rm abs}$}
\def\zem{$z_{\rm em}$}

\def\21{21-cm}

\def\t0{T$_{0}$}

\def\c21{$C_{21}$}

\def\J16{J$1621+0758$}

\title[UFOs in J1621+0758]{Correlated time variability of multi-component high velocity outflows in J162122.54+075808.4\thanks{Based on observations collected at Southern African Large Telescope (SALT; Programme IDs 2015-1-SCI-005, 2018-1-SCI-009, 2019-1-SCI-019 and 2020-1-SCI-011) and the European Organisation for Astronomical Research in the Southern Hemisphere under ESO programme 093.A-0255.}}

\author[Aromal et al.]{
P. Aromal$^{1}$\thanks{E-mail: aromal@iucaa.in (PA)},
R. Srianand$^{1}$,
and P. Petitjean$^{2}$
\\
$^{1}$IUCAA, Postbag 4, Ganeshkind, Pune 411007, India\\
$^{2}$ Institut d'Astrophysique de Paris, Sorbonne Universit\'es and CNRS, 98bis boulevard Arago, 75014 Paris, France\\
}

\date{Accepted XXX. Received YYY; in original form ZZZ}

\pubyear{2015}

\begin{document}
\label{firstpage}
\pagerange{\pageref{firstpage}--\pageref{lastpage}}
\maketitle

\begin{abstract}
We present a detailed analysis of time variability of two distinct \civ\ broad absorption line (BAL) components seen in the spectrum of J162122.54+075808.4 (\zem = 2.1394) using observations from SDSS, NTT and SALT taken at seven different epochs spanning about 15 years.
The blue-BAL component (with an ejection velocity, $v_{\rm e}\sim37,500$ \kms) is an emerging absorption that shows equivalent width variations and kinematic shifts consistent with acceleration. The red-BAL component ($v_{\rm e} \sim 15,400$ \kms) is 
a three component absorption. One of the components
is emerging and subsequently disappearing. The two 
other components show
kinematic shifts consistent with acceleration coupled with equivalent width variability.  Interestingly, we find the kinematic shifts and equivalent width variability of the blue- and red-BAL components to be correlated. 
While the \civ\ emission line flux varies by more than 17\% during our monitoring period, the available light-curves (covering rest frame 1300-2300\AA) do not show more than a 0.1~mag variability in the continuum. This  suggests that the variations in the ionizing flux are larger than that of the near-UV flux. However, the correlated variability seen between different BAL components cannot be explained solely by photoionization models without structural changes.  In the framework of disk wind models, any changes in the radial profiles of density and/or velocity triggered either by disk instabilities or by changes in the ionizing radiation can explain our observations.
%
High resolution spectroscopic monitoring of \J16 is important to understand the physical conditions of the absorbing gas and thereby to constrain the parameters of disk-wind models. 
\end{abstract}




\section{Introduction}

Strong outflows with velocities reaching up to few 10,000 \kms\ are seen in quasar spectra as blue-shifted broad absorption lines (BALs) with velocity widths of several $1000$~km~s$^{-1}$. Associated quasars are commonly known as `BAL' quasars \citep{Weymann1991}. 
These powerful outflows can carry a significant amount of mechanical energy and momentum which can in principle contribute to AGN feedback mechanisms, regulating the central black hole growth and the host galaxy evolution \citep{ostriker2010,kormendy2013}. They might interact with the inter-stellar medium of the host galaxies which can result in quenching or enhancement of star formation. If these outflows can escape the galactic potential wells they can also contribute to the chemical enrichment of the intergalactic medium (IGM) in the vicinity of galaxies.

The time variability of BAL profiles is a powerful tool for studying the origin and evolution of BAL outflows. 
Monitoring BAL variability over different time scales can put tight constraints on the BAL lifetime, location of the outflow etc., and provide significant insights on the origin and physical mechanisms driving the flow.
The time variability of \civ, Si~{\sc iv} and Mg~{\sc ii} BAL profiles have been studied extensively in the literature.
BAL variability includes extreme optical depth variations like emergence, disappearance and kinematic shift of BALs \citep{Filiz2013,rogerson2018,vivek2018,cicco2018,mcgraw2017}. Possible reasons for these variations include : (i) fluctuations in quasar ionizing flux, (ii) changes in covering fraction of the outflow with respect to the central source, (iii) bulk motion of the outflow. It has been found that significant changes in the equivalent width and/or shape of the BAL trough occur typically on time scales of few months to years in the quasar rest frame \citep[see for example,][]{Filiz2013,Vivek2014}.

Among all outflows, ultra-fast AGN outflows (UFOs) are of great interest as they probe the inner regions of 
the central engine. 
\citet{Tombesi2010} defined UFOs as highly ionized absorbers detected mostly through Fe K shell absorption lines in X-rays at velocities $v_{\rm outflow} \ge 10^4$~\kms. The high velocities and high ionization state of the absorbers (mainly H-like and He-like Fe) 
are usually attributed to their origin from within a few hundred gravitational radii from the central black hole where high energy photons are produced. 
%
UV absorptions from such BALs are commonly observed
\citep[][]{Srianand2001}. 
They tend to show larger variability, emergence and acceleration in their BAL profiles \citep{capellupo2011, grier2016, vivek2018}. 
Recently, \citet{Rodriguez2020} have studied a sample of extremely high velocity outflows (EHVO, defined as absorption with outflow velocities between 0.1 and 0.2c).
These outflows are detected in 0.6\% of quasars searched for \civ\ absorption. While \lya\ associated with the \civ\ component is either not detected or weak, 
%
$\sim$50\% of these BALs show associated \nv\ absorption and about 13\% show O~{\sc vi} absorption. These objects tend to have higher black hole mass and bolometric luminosities compared to the general population.
Unfortunately, X-ray observations of these high-$z$ EVHOs are not available which makes it difficult to establish any connection between them and X-ray detected UFOs.

Possible acceleration mechanisms through which outflows reach such high velocities are still debated. Photon scattering by free electrons alone is not sufficient for the absorbing gas to be accelerated to UFO (or EHVO) velocities. Radiative acceleration through UV line absorption 
has often been invoked \citep{arav1994, murray1995,proga2000} since observational evidences such as line-locking \citep{srianand2000} or Ly-$\alpha$ ghost \citep{arav1996} strongly support line-driven BAL outflows. Another alternate mechanism is known as magnetic driving which can also explain highly ionized UFOs \citep{dekool1995}. Even though these models are successful in explaining the observed BAL features, they all predict the outflows to be located very close to the central source at a distance of the order of 0.01-0.1 pc. If true, despite carrying large amounts of mechanical energy and momentum, they may have little
influence on the 
large-scale star formation properties of the host galaxy.

The SDSS survey provides us with multi-epoch spectroscopic observations of BAL quasars 
allowing us to study BAL variability over the period the survey is performed. We have identified a sample of 31 BAL quasars showing large variability in the SDSS and conducted a spectroscopic monitoring at shorter time scale using the South African Large Telescope (SALT). 
Here, we present a detailed  analysis of an interesting BAL quasar, J162122.54+075808.4 ($z_{\rm em} = 2.1394$), hereafter refer to as \J16, from our sample. 
This quasar shows emergence of a BAL component with UFO (and EHVO) velocities followed by a kinematic shift in the BAL profile implying acceleration along our line of sight. 
In addition to the emerged component, an already existing BAL at comparatively lower velocities (while still satisfying the definition of UFOs) also shows high variability. Interestingly, the variability of these two distinct components are highly correlated.
We also report time variability of \civ\ and \siv\ emission line equivalent widths during our monitoring period. We find that the available light curves are consistent with no long-term brightening or fading of \J16. 
{\it Thus \J16 is an interesting target to understand the origin and evolution of UFOs (or EHVOs) through variability studies. This forms the main motivation of this paper.}

The paper is organized as follows. In section~\ref{sec:obs} we present the details of spectroscopic observations of \J16\ using SALT and ESO/NTT
and data reduction. 
In section~\ref{sec:abs}, we provide the details of the \civ\ BAL components and quantify their rest equivalent width and kinematic variability. We probe the possible correlated variability in the \civ\ rest equivalent width and outflow velocities between different BAL components. In section~\ref{sec:continuum}, we study the long-term rest UV continuum flux and colour variability of \J16\ using all the available spectra and photometric lightcurves.
In section~\ref{sec:discuss}, we discuss our results in the framework of simple photoionization models and simple disk-wind model predictions.
%
We summarize  our main findings in section~\ref{sec:summary}.

\section{Observations $\&$  Data reduction}
\label{sec:obs}
\begin{table*}
\begin{threeparttable}
    \centering
\caption{Log of observations and details of spectra obtained at different epochs 
}
 \begin{tabular}{ccccccccccc}
  \hline
   Telescope & \multicolumn{2}{c}{Date \tnote{a}} & Exposure  & Spectral  & S/N \tnote{b} & $EW_{1548}^B$ \tnote{c} & $EW_{1548}^R$ \tnote{d} & $A_V$ \tnote{e} & $p$ \tnote{f} & $g-r$ \tnote{g} \\ 
       used  & (D/M/Y) & (MJD) &  time (s)& res. (\kms) & &(\AA) & (\AA) \\
  \hline\hline
  SDSS & 11-05-2005 & 53501 & 4200 & 150 & 19.41 & - & $2.70 \pm 0.17$ & 0.03 & $0.16 \pm 0.01$ &.... \\ 
  SDSS & 30-03-2012 & 56016  & 4200 & 150 & 28.69 & $3.66 \pm 0.09$ & $4.72 \pm 0.12$ & 0.17 & $0.59 \pm 0.01$ & ....\\
  NTT & 21-07-2014 & 56859 & 2 $\times$ 1800 & 650 & 7.11 & $5.11 \pm 0.68$ & $8.63 \pm 0.79$&.... & ....& ....\\ 
  SALT & 08-06-2015 & 57181  & 2000 & 304 & 40.35 & $5.52 \pm 0.08$ & $6.99 \pm 0.09$ & 0.11 & $0.43 \pm 0.01$ & ....\\ 
  SALT & 07-05-2018 & 58245  & 2 $\times$ 1150 & 304 & 18.29 & $3.39 \pm 0.19$ & $5.46 \pm0.25$ & 0.09 & $0.36 \pm 0.01$ & $0.22 \pm 0.04$\\
  SALT & 03-05-2019 & 58606  & 2 $\times$ 1150 & 304 & 26.91 & $2.69 \pm 0.12$ & $4.76 \pm 0.15$ & 0.07 & $0.28 \pm 0.01$ & $0.22 \pm 0.04$\\
  SALT & 21-05-2020 & 58991  & 2170 & 304 & 44.86 & $3.03 \pm 0.07$ & $4.01 \pm 0.09$ & 0.16 & $0.65 \pm 0.01$ & $0.27 \pm 0.05$\\    
  \hline
 \end{tabular}
 \begin{tablenotes}\footnotesize
    \item[a] Date of observations.
    \item[b] Signal-to-noise ratio per-pixel calculated over the wavelength range 5200-5500 \AA~ for SDSS and SALT data and 4900-5180 \AA~ for NTT data.
    \item[c,d] Total \civ\ rest equivalent width of "blue-" and "red-BAL" components obtained by integrating over the absorption profile respectively.
    \item[e] Best fit parameter value of $A_{\rm V}$ as obtained from $\chi^2$ minimization.
    \item[f] Best fit parameter value of $p$ in  $C (\frac{\lambda}{\lambda_0})^p$, mimicing a change in the power-law spectrum relative to the template, as obtained from $\chi^2$ minimization.
    \item[g] g-r color obtained from the closest (i.e within 1 day) available photometric measurements as explained in \ref{sec:lc}. 
 \end{tablenotes}
\label{tab_obs}
\end{threeparttable}
\end{table*}


A \civ\ broad absorption with \zabs $\sim$1.9818 is detected in the initial spectrum of \J16 obtained for SDSS in the year 2005. A new \civ\ BAL at \zabs $\sim$ 1.7687 emerged in another SDSS spectrum taken in year 2012. 
This prompted us to undertake spectroscopic monitoring of \J16 using the Southern African Large Telescope \citep[SALT,][]{buckley2005}. 

In the SDSS catalog \citep{paris2012}
the quoted emission redshift of \J16 is \zem = $2.13449\pm0.00032$. 
\citet{Hewett2010} derive a systemic redshift of \zem = $2.139449\pm0.000335$ from the fit of the C~{\sc iii}] emission. We use this latter value as the systemic redshift for all discussions in this paper. Note that given the high velocities of the BALs we observe here, the exact emission
redshift has little influence on our study.
Indeed, with respect to this systemic redshift the above identified two BAL components have ejection velocities of $\sim$ 15,400 \kms\ and  37,500 \kms\ based on the positions of the maximum \civ\ optical depths.
We denote the two BAL components as ``red-" and ``blue-" BALs respectively

\citet{hidalgo2020} found that, in a parent sample of 6760 quasars (please refer to section 2 of their paper for details) from \citet{paris2012}, only $\sim$14$\%$ show \civ\ BALs according to the conventional definition \citep{Weymann1991}
and an even lesser fraction of $\sim$0.6 $\%$ show extremely high velocity outflow (EHVO) with velocities ranging from 30,000 to 60,000 kms$^{-1}$. \J16 falls into this special type of BAL quasars due to the high ejection velocity of one of the \civ\ absorption components.

The log of SDSS, NTT and SALT observations, details of the spectra, absorption lines and reddening measurements are summarized in Table~\ref{tab_obs}.
New spectroscopic observations of \J16 were carried out as part of our BAL quasar spectroscopic monitoring program with the SALT   (Program IDs : 2015-1-SCI-005, 2018-1-SCI-009, 2019-1-SCI-019 and 2020-1-SCI-011). 
We used the Robert Stobie Spectrograph \citep[RSS,][]{Burgh2003,Kobulnicky2003} in the long-slit mode using a slit 1.5" wide and the PG0900 grating (kept at a position angle, PA = 0.0). With GR angle=$14.75^{\circ}$ and CAM angle=$29.5^{\circ}$, this setting provides a wavelength coverage of $4060-7120$ {\AA} excluding the 5065-5122 {\AA} and 6115-6170 {\AA} regions falling in the CCD gaps. The spectral resolution and average signal-to-noise ratio (S/N) obtained are summarized in columns 5 and 6 of Table~\ref{tab_obs}. Our SALT spectra typically have a spectral resolution of $R \sim 985$ (roughly a factor of two smaller than that of SDSS) at central wavelength $5620$ \AA\ and S/N in the range 18-45  per-pixel.

The raw CCD frames have been preliminary processed 
using the SALT data reduction pipeline \citep{Crawford2010}. We used then standard IRAF\footnote{IRAF is distributed by the National Optical Astronomy Observatories, which are operated by the Association of Universities for Research in Astronomy, Inc., under cooperative agreement with the National Science Foundation.} procedures to reduce the resulting 2D spectra. Flat-field corrections and cosmic ray zapping were applied to all science frames. We extracted the one dimensional quasar spectrum from the background subtracted 2D science frames from each epoch using the IRAF task ``Apall". Wavelength calibration was performed using standard Argon lamp spectra. 
In addition, skylines from the wavelength calibrated spectrum were matched with the sky line atlas provided by SALT and, if needed, corrections were applied to increase wavelength accuracy. Similarly, flux calibration was performed using reference stars (G93-48 and LTT4364) observed close to our observing nights. 

Prior to our SALT observations we also observed \J16 using NTT as a backup target (for the ESO program 093.A-0255; PI: R. Srianand) 
under overcast conditions, 
high airmass ($\sim 2$) and good seeing (i.e $\sim$1") conditions. We have used Grism GR\#7 of EFOSC2 with a slit width of 1.5". We obtained two exposures of 30 min each and the data were recorded with a 2x2 CCD binning. The data were reduced using standard IRAF procedures. As we do not have standard star observations, flux calibration was not performed. Also the available arc lamp spectrum does not have strong emission lines close to the blue-BAL component. Therefore, we correct the wavelength solution 
using nearby intervening absorption lines. The details of the NTT spectra  are also summarized in Table~\ref{tab_obs}.

We detect intervening Mg~{\sc ii} absorption systems at \zabs = 1.83739 and 1.84201 and C~{\sc iv} systems at \zabs = 1.89483, 2.00575 and 2.04649. Various narrow metal absorption lines from these systems are present in the same wavelength range 
as the ``red-" and ``blue-BAL" components. These lines give us strong
constraints
on any velocity shift in the \civ\ broad absorption as a function of time.  We believe these are intervening absorbers as, (i) the ejection velocities are larger than  5000 \kms, (ii) we do  not detect associated N~{\sc v} absorption and (iii)  \civ\ equivalent width does not show time variations between two SDSS epochs where the \civ\ doublet is resolved.
%
In total, we have spectroscopic observations of \J16\ during 7 epochs spread over 15 years. We denote them by epoch-1, epoch-2, etc., in chronological order. 

\section{Analysis of spectral variability}
\label{sec:abs}

\begin{figure*}
\centering
\includegraphics[viewport=105 25 1220 650,scale=0.45,clip=true]{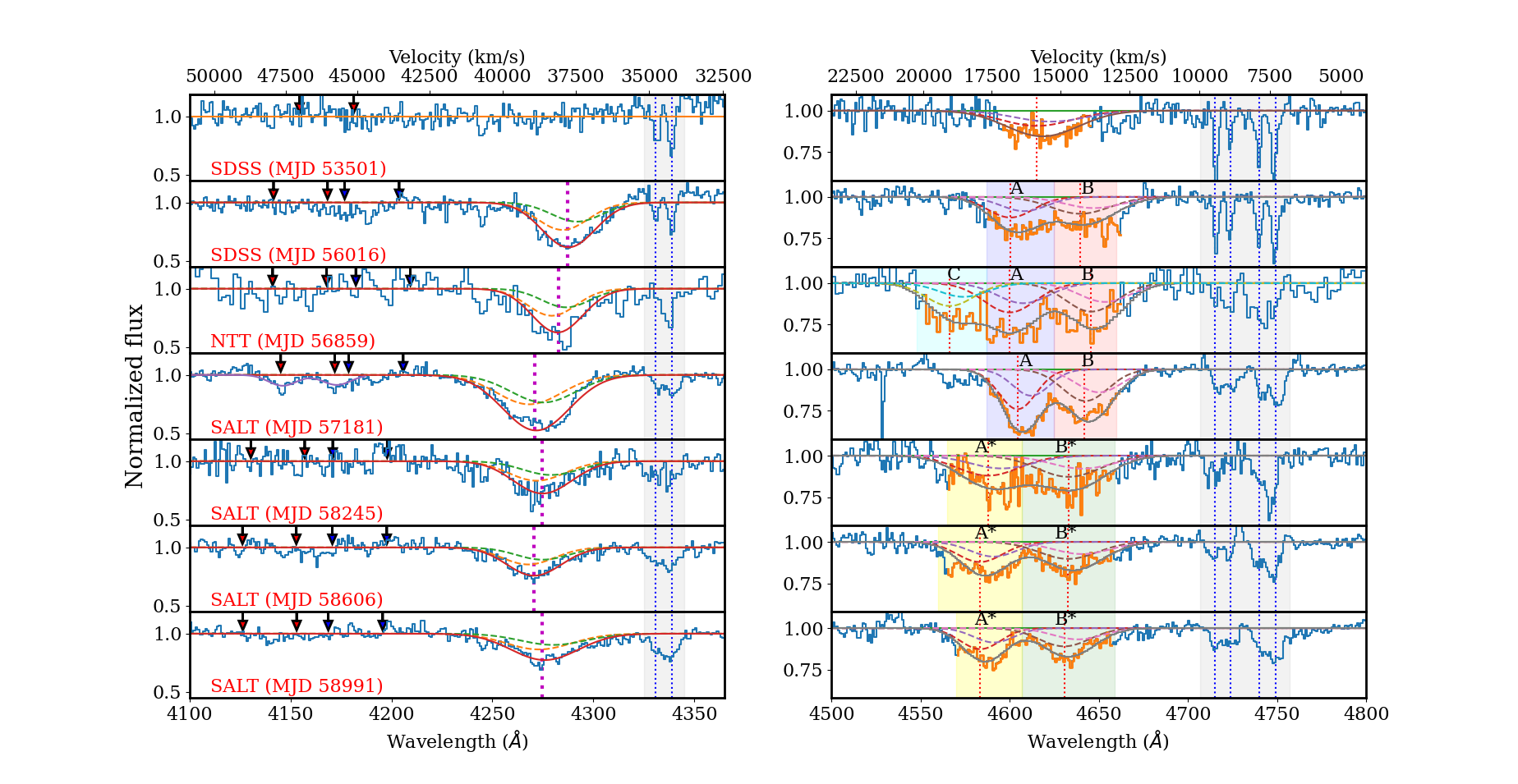}
\caption{Normalized spectra (using lower order polynomials) showing the time evolution of ``blue" (left panels) and ``red" (right panels) \civ\ BALs. 
The relative velocity scale with respect to the systemic redshift of \J16 (i.e \zem = 2.13945)
is given at the top. The absorption profiles are fitted
with gaussian components.
 The best fits 
are over-plotted as red and gray solid lines in the left and right panels respectively. In the right panel, the filled regions marked with A, B and C correspond to the regions where components A, B and C of the red-BAL are located. The dashed lines in the figure correspond to the centers of individual gaussian components contributing to the best fit (representing the \civ\ doublet lines for each BAL component).   Intervening absorption lines are marked by blue dotted vertical lines inside the gray shaded regions. Black vertical arrows in the left panels indicate the expected positions of the Si~{\sc iv} absorption 
associated to the \civ\ components identified in the red BAL shown in the corresponding right panels.}
\label{fig:absvar}
\end{figure*}

 In this section, we present the observed properties of the two BAL components, their time evolution and correlated variabilities. We also study in detail the variability of the emission line profiles.
 For absorption line studies, we have normalized the spectra with the unabsorbed continuum determined
 by fitting low order polynomials to the quasar emission. 
The normalised spectra around the \civ\ absorption are shown in Fig.~\ref{fig:absvar}. It is most likely that all absorptions we see are part of the same flow, for presentation purpose we consider "blue-" and "red-BAL" components separately in our discussion.

\subsection{Kinematic and  equivalent width variability of the red-BAL}
\label{sec:red_bal}

As seen from the right panels of Fig.~\ref{fig:absvar}, the \civ\ absorption profile of the red-BAL shows significant variability in terms of velocity spread, equivalent width, number of distinct absorption components and kinematic shifts. 
To quantify these,
we approximated the \civ\ absorption profile of the red-BAL component with multiple Gaussians (each velocity component will have a double Gaussian to take care of the \civ\ doublet absorption). These fits are also shown in Fig.~\ref{fig:absvar} and corresponding fit parameters are summarized in Table~\ref{tab_gauss}. Column 3 of this table provides the velocity centroid of the individual Gaussian components. The difference in velocity centroid between two subsequent epochs is used to measure the acceleration given in column 4. The Gaussian velocity widths quoted in column 5 of this table for individual components are consistent with them being mini-BALs.  
The equivalent widths quoted in column 6 is the sum of the \civ~$\lambda1548$ and \civ~$\lambda1550$ doublet lines and the errors are computed by fitting 1000 randomly generated spectra using the measured flux errors  
to randomize the flux in each pixel.
The associated Si~{\sc iv} absorption is detected in component A during epoch-4. We were able to fit both \civ\ and Si~{\sc iv} with a consistent set of gaussian parameters (i.e width and centroid) during this epoch.
The rest equivalent widths (or 3$\sigma$ upper limits) of \civ\ and Si~{\sc iv} are provided in columns 6 and 7 respectively.

During the first SDSS epoch (epoch-1), we identify the red-BAL as a single shallow and broad (FWHM$\sim$3050 \kms) absorption centered around outflow velocity of 15,000 \kms\ and with a \civ\ equivalent width of $2.50\pm0.29$ \AA. 
We do not detect any other associated absorption, neither Si~{\sc iv} nor Al~{\sc iii}. This epoch spectrum does not cover the expected wavelength range of N~{\sc v} and \lya\ absorptions.
%

During epoch-2, the \civ\ BAL becomes stronger, the velocity range covered by the absorption has increased (i.e $\sim$5000 \kms) and the absorption profile also splits into two distinct components (separated by 2532.0$\pm$210.8 \kms) each having similar \civ\ equivalent widths (i.e 2.14$\pm$0.32 and 2.24$\pm$0.32~\AA\ respectively; see Table~\ref{tab_gauss}). It is also evident that \civ\ absorption profile for both the components are deeper than what is seen in the first epoch spectrum.
We denote these two components as A and B in Fig~\ref{fig:absvar}. 
The Si~{\sc iv} absorption from these components is not clearly detected at this epoch. The SDSS spectrum obtained during this epoch
also covers the expected wavelength range of \lya\ and N~{\sc v} absorption from the red-BAL. 
We do not find any strong absorption line at the expected position of H~{\sc i} \lya\ absorption. 
However, we cannot rule out a possible weak absorption feature.
We fit two Gaussians corresponding to A and B components of the red-BAL and derive upper limits on the \lya\ equivalent widths to be 1.80 \AA\ and 1.10 \AA\ respectively. If we assume the absorption to be on the linear part of the curve of growth, these limiting equivalent widths correspond to 
H~{\sc i} column density of $N$(H~{\sc i}) $\le 1.7\times 10^{15}$ cm$^{-2}$ and $N$(H~{\sc i}) $\le 1.1\times 10^{15}$ cm$^{-2}$ for A and B respectively. The actual limiting column densities can be higher if the absorbing gas does not cover the background source completely.
%
%

In the spectra obtained using NTT (epoch-3) the \civ\ absorption, in addition to showing A and B components albeit with enhanced absorption strength, also shows
a newly emerging strong component at $\sim$18,500 \kms (denoted as component C). This new component has a velocity separation of $2156\pm239$ \kms  with respect to component A which is very close to the Si~{\sc iv} doublet splitting of 1925 \kms.
High resolution spectra would have been important to confirm such a velocity coincidence and support the line driven acceleration.
This absorption is clearly seen in both the spectra obtained with NTT with a \civ\ rest equivalent width of 1.95$\pm$0.33\AA. 
As before we do not detect Si~{\sc iv} absorption associated to any of these three \civ\ components in our spectrum at this epoch.
Our NTT spectrum also covers \lya\ and N~{\sc v} regions but we do not detect any significant absorption at the expected positions.
{\it Our observation constrains the emerging time-scale for component "C" to $\leq$ 268.5 days in the quasar rest frame.} 


In the spectra taken during our first SALT epoch (epoch-4) component C, which was found to be emerging in the previous epoch, has become very weak (with a 3$\sigma$ upper limit of W(\civ) $\leq 0.38$ \AA). 
Thus the overall  time-scale over which this component emerged and subsequently disappeared along our line of sight is $\le 424.4$ days in the quasar frame. 
%
This time scale is within the smallest observed so far \citep{filiz2012, mcgraw2017, cicco2018}.
%
%
Components A and B are seen nearly at the same redshifts as in epoch-3 and their
equivalent widths reach their maximum values (i.e 3.29$\pm$0.04 and 3.32$\pm$0.06\AA~ respectively for A and B) 
(see Table~\ref{tab_gauss}). 
The optical depth variations are non-uniform throughout the BAL profile component A showing larger absorption optical depth changes compared to B (see Fig.~\ref{fig:absvar}). Interestingly, we observe the resolved Si~{\sc iv} absorption associated with component A
(see the absorptions at the location of the first two arrows from the left in the corresponding left panel for epoch 4
of Fig.~\ref{fig:absvar}). However, we do not detect Si~{\sc iv} from component B. Unfortunately during this epoch our SALT spectrum does not cover the \lya\ and N~{\sc v} absorption wavelength range. 

In the next SALT epoch (epoch-5), the BAL becomes weaker with significant (non-uniform) changes in the absorption profile, indicating a possible kinematic shift. 
%
We note that the wavelength calibration is ascertained by the presence 
of intervening absorption lines at the same position in all spectra. These intervening absorption are shown with vertical dotted lines in gray shaded regions of Fig.~\ref{fig:absvar}. The absorption lines seen around 4720\AA\ are \civ\ doublets from the \zabs=2.04649 system. 
Lines seen around 4750 \AA\ are
\alii\ absorption from the \mgii\ systems at \zabs = 1.83739 and \zabs = 1.84201. The \siii\ absorption from the same \mgii\ systems (around 4340 \AA) are shown in the left panel. We also confirm the wavelength solutions using the wavelength of the sky lines.

The \civ\ absorption profile at this epoch is also well fitted with two gaussian components. We call these components as ``A*" and ``B*".
At the same time the Si~{\sc iv} absorption associated with component A at epoch-4, has disappeared.  
The variations in the profile seen at this epoch with respect to the previous epochs are consistent with two possibilities (i) the \civ\ components A and B have accelerated
(with associated variations in \civ\ optical depth) 
or (ii) the components A and B disappeared along our line of sight and new components A* and B* have emerged close to (albeit at slightly higher velocities) compared to A and B. 
The presence of two components with similar velocity separation favors the first possibility. However the optical depth change and the non-detection of Si~{\sc iv} absorption will be inconsistent with  simple acceleration without profile changes discussed in the literature \citep{grier2016}.
While we will consider the above two possibilities, for simplicity we quote these
two components as A and B from now on.
%
%

During epoch-6 (i.e., MJD 58606) the two \civ\ absorption components A and B 
are well defined and there is a slight reduction in the \civ\ equivalent width with a possible kinematic shift (see Table~\ref{tab_gauss}).
During this epoch we do not detect any Si~{\sc iv} absorption. During epoch-7 (i.e., MJD 58991) \civ\ rest equivalent widths of A and B have further reduced without any additional kinematic shift.

It is also clear, from the 3$\sigma$ upper limits on Si~{\sc iv} equivalent widths for component A, that not only the C~{\sc iv} equivalent width has changed but the ratio of Si~{\sc iv} to C~{\sc iv} equivalent widths has also changed with time (see Table~\ref{tab_gauss}). 

\subsection{Emergence, kinematic and equivalent width variations of the blue-BAL}
\label{sec:blue_bal}
\begin{figure}
\centering
\includegraphics[viewport=50 5 1200 620,scale=0.22,clip=true]{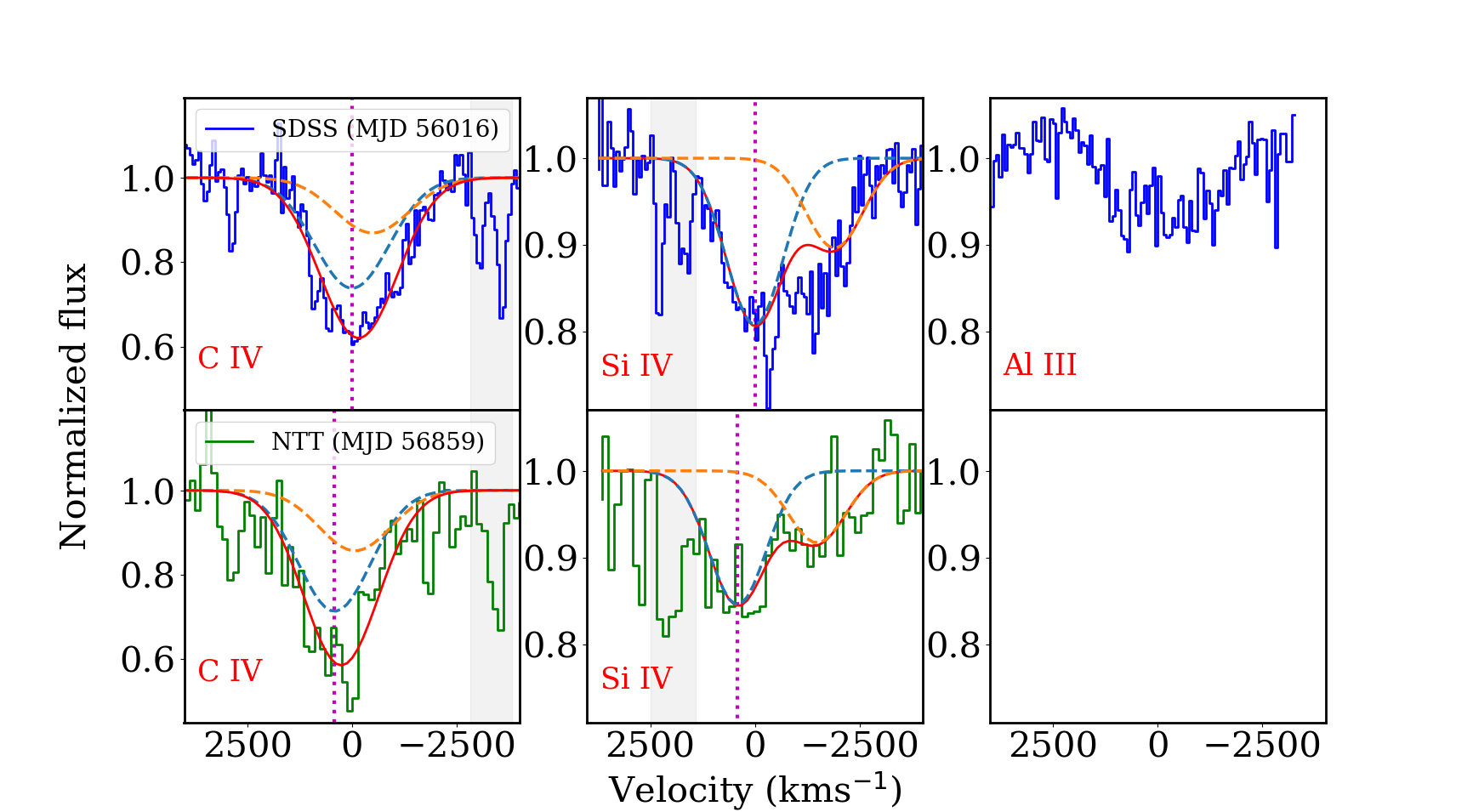}
\caption{Gaussian fits to the C~{\sc iv} and  Si~{\sc iv}   absorption from the blue-BAL component (red solid lines) seen in spectra obtained with
SDSS (MJD 56016; Top panel) and NTT (MJD 56859: bottom panel) epochs. The dashed lines in the figure correspond to the individual gaussian components contributing to the best fit (representing the \civ\ and \siv\ doublet lines for each BAL component). Vertical dotted lines show the velocity centroid of the gaussian component corresponding to the  blue member of the \civ\ (in the left panel) and \siv\ (in the center panel) doublet lines with respect to the absorption redshift \zabs = 1.7688 ( measured using optical depth weighted centroid). The Al~{\sc iii} absorption detected during epoch-2 is shown in the top right panel.}

\label{fig:bluesi4}
\end{figure}

Emergence of the blue-BAL, located on the blue side of the Si~{\sc iv} emission line, is apparent in the epoch-2 SDSS spectrum with an outflow velocity of $\sim$37,000 \kms. The emergence time-scale of this component is $\le$802 days in the quasar's rest-frame.
We also detect the Si~{\sc iv} doublet associated with this \civ\ absorption on top of the quasar \lya+N~{\sc v} emission line. These absorption lines are well fitted with a single Gaussian component (see Figs~\ref{fig:absvar} and \ref{fig:bluesi4}). The Gaussian parameters for both \civ\ and Si~{\sc iv} absorptions are summarised in Table~\ref{tab_gauss}. We also detect an absorption at the location of Al~{\sc iii} (see Fig.~\ref{fig:bluesi4}) with a rest equivalent width of 0.93$\pm$0.10 \AA.

In the epoch-3 spectrum obtained with NTT, we observe
a shift in the \civ\ absorption feature without appreciable change in its rest equivalent width. This possible acceleration signature is illustrated by the Gaussian fits to the \civ\ and Si~{\sc iv} absorptions shown in Fig~\ref{fig:bluesi4} showing that the kinematical shifts are consistent for the \civ\ and Si~{\sc iv} profiles. 
The acceleration inferred from the Gaussian centroid of the \civ\ absorption (see Table~\ref{tab_gauss}) is not statistically significant however due to poor SNR of our NTT data. 
Unlike the \civ\ absorption, the Si~{\sc iv} absorption shows approximately a factor of two reduction in the rest equivalent width.

As for the red-BAL, the \civ\ equivalent width of the blue-BAL reaches its maximum in the epoch-4 SALT spectrum (on MJD 57181). 
From Fig.~\ref{fig:absvar}, it is clear that the noticed change in the \civ\ absorption profile is consistent with acceleration. 
If we consider the Gaussian centroids obtained from epoch-3 and 4 spectra we get an acceleration of $+8.84\pm3.61$ cm s$^{-2}$ (significant at the 2.4$\sigma$ level) 
Instead, if we consider epochs 2 and 4 the inferred acceleration is $3.57\pm0.08$ cm s$^{-2}$. This is one of the largest accelerations inferred among BAL quasars \citep[see figure 6 of][]{grier2016}.
Our SALT spectra of this epoch and subsequent epochs do not cover the Si~{\sc iv} region.
%

From Fig.~\ref{fig:absvar} (also see Table~\ref{tab_gauss}) we find that for the next three epochs the strength of the \civ\ absorption has reduced as in the case of the red-BAL components. We also notice the changes in the centroid of the Gaussian components used to fit the \civ\ absorption. However these shifts (seen after the epoch-4) are not
monotonous and their significance is typically at the 2-3$\sigma$ level (see Table~\ref{tab_gauss}). 
%
%

\begin{figure}
\hspace*{-0.5cm}
\centering
\includegraphics[viewport=30 20 1000 600,scale=0.32,clip=true]{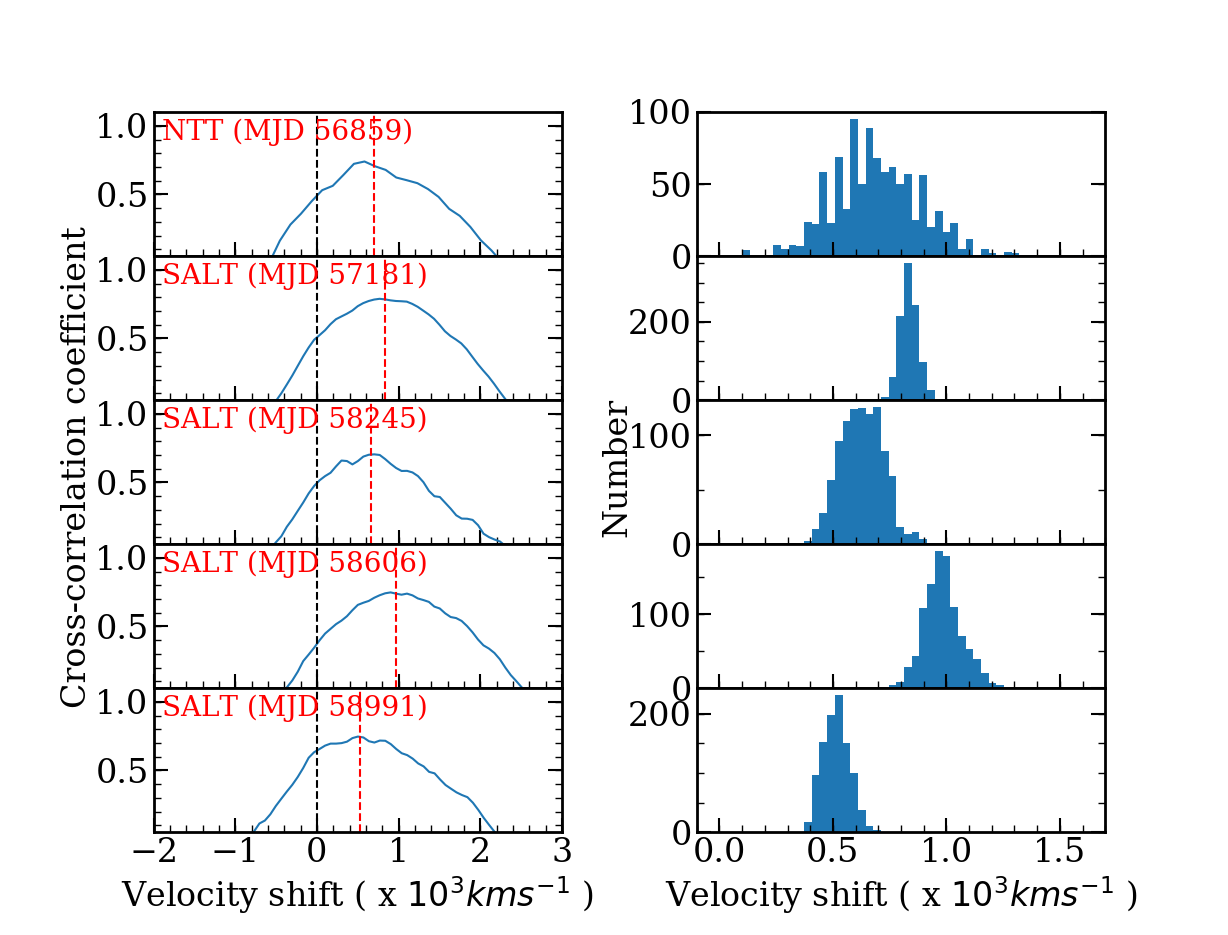}
\caption{{\it Left Panels:} Cross-correlation function (CCF) for the \civ\ blue-BAL component obtained at different epochs with respect to the \civ\ profile observed in epoch-2.The centroid shift of the BAL profile is shown by the red dashed line and the zero velocity shift is shown by the black dashed line. {\it Right panel:}
%
Corresponding cross-correlation centroid distributions (CCCD) from Monte Carlo simulations. }
\label{fig:ccf}
\end{figure}

The kinematic shifts are usually quantified using cross-correlation analysis \citep[see for example,][]{grier2016,Joshi2019}. 
We thus calculate the cross correlation between the BAL profile in the epoch-2 spectrum with those obtained at subsequent epochs.
We estimate the shift in wavelength using the 
location of the peak in the cross-correlation coefficient ($r_{\rm peak}$) 
estimated using only points around the peak with values greater than 0.8$r_{\rm peak}$. Taking into account measurement uncertainties, we randomize the flux in each pixels of both spectra
and generate 1000 realizations to measure the CCF and in each case the corresponding peak and centroid.
The final velocity shift is the median of the cross-correlation centroid distribution (CCCD) and the 1 $\sigma$ uncertainty is the central interval encompassing 68 percent of CCCD. The measured CCFs for different epochs are summarised in Fig~\ref{fig:ccf} and Table~\ref{tab_ccor}.

It is clear from this figure that compared to the epoch-2 spectrum the blue-BAL
component observed in subsequent epochs shows clear sign of increase in the ejection velocity by $\sim$530 to $\sim$975 \kms. 
The measured acceleration between different epochs varies from 0.64 cm s$^{-2}$ to 2.81 cm s$^{-2}$ (see Table~\ref{tab_ccor}). 
These values are inconsistent with a nearly constant velocity one expects from simple disk wind models in late stages of the flow. Note the BAL quasars showing acceleration signatures often tend to show such behavior whenever more than two epoch observations are available
\citep{grier2016,Joshi2019}. The blue-BAL that is seen as a single component in our low resolution spectrum may have a multiple absorbing component structure. Any change (either individual optical depths or velocity centroids) in these components can lead to small centroid shifts in the overall absorption. While this cannot explain the large shifts seen in the earlier epochs, it could contribute to smaller shifts we see in the later epochs.
%
 The probability that the profile shapes between two epochs are identical after applying appropriate kinematic shifts as given in Table~\ref{tab_ccor} is nearly zero for all the epochs considered in the cross-correlation analysis which is
consistent with the appreciable profile variability we observe (Fig.~\ref{fig:absvar}).  This suggests that the blue-BAL in addition to showing appreciable velocity shift also has strong optical depth variability. Thus the variability found is inconsistent with a single absorbing gas accelerating along our line of sight without any change in the strength of the absorption it produces.

\begin{table*}
\begin{threeparttable}
    \centering
\caption{Parameter values obtained from gaussian fits of the BAL components. }

\begin{tabular}{ccccccc}
  \hline
   BAL component & Epoch \tnote{a} & Velocity centroid ($v_c$) \tnote{b} & Acceleration  & Velocity width \tnote{c}  &  EW (C~{\sc iv}) \tnote{d} & EW (Si~{\sc iv}) \tnote{e} \\ 
         &  &  (\kms) & (cm s$^{-2}$) &(\kms) & (\AA) & (\AA)\\
  \hline\hline
   & 1 & - & - & - & - & - \\
   & 2 & $16478 \pm 159$ & - & $787 \pm 77$ & $2.14 \pm 0.31$ & $\le 0.30$ \\
   & 3 & $16588 \pm 150$ & $+0.47 \pm 0.94$ & $789 \pm 60$ & $3.35 \pm 0.53$ & $\le 1.83$ \\  
   Red BAL (A) & 4 & $16211 \pm 60$ & $-4.25 \pm 1.82$  & $669 \pm 24$ & $3.29 \pm 0.04$ & $2.04 \pm 0.09$   \\  
   & 5 & $17257 \pm 199$ & $+3.57 \pm 0.71$   & $1165 \pm 624$ & $2.70 \pm 1.33$ & $\le 0.71$  \\
   & 6 & $17574 \pm 101$ & $+3.19 \pm 2.25$    & $810 \pm 84$ & $2.18 \pm 0.18$ & $\le 0.38$  \\
   & 7 & $17571 \pm 59$ & $ -0.03 \pm 1.11$    & $693 \pm 31$ & $1.93 \pm 0.07$ & $\le 0.21$   \\
   \hline
   & 1 & - & - & - & - & - \\
   & 2 & $13946 \pm 291$ & -  & $1005 \pm 154$ & $2.24 \pm 0.32$ & $\le 0.33$  \\
   & 3 & $13653 \pm 175$ &  $-1.26 \pm 1.46$  & $936 \pm 135$ & $2.52 \pm 0.55$ &  $\le 1.92$ \\  
   Red BAL (B) & 4 & $13792 \pm 40$ & $+1.57 \pm 2.02$  & $781 \pm 17$ & $3.32 \pm 0.06$ & $\le 0.35$   \\  
   & 5 & $14330 \pm 185$ & $ +1.83 \pm 0.64$ & $1328 \pm 488$ & $3.19 \pm 1.20$ & $\le 0.68$   \\
   & 6 & $14386 \pm 101$ & $ +0.57 \pm 2.11$ & $987 \pm 92$ & $2.22 \pm 0.16$ & $\le 0.40$  \\
   & 7 & $14519 \pm 55$ & $+1.25 \pm 1.08$   & $881 \pm 62$ & $2.02 \pm 0.10$ & $\le 0.23$  \\
   \hline
  Red BAL (C) & 3 & $18765 \pm 177$ & - & $799 \pm 59$ & $1.95 \pm 0.33$ & - \\
   \hline  
   & 1 & - & - & - & - & - \\
   & 2 & $37592 \pm 39$ & - & $783 \pm 21$ & $4.04 \pm 0.09$ & $2.09 \pm 0.17$\\
   & 3 & $37951 \pm 320$  & $+1.54 \pm 1.39$  & $807 \pm 254$ & $4.36 \pm 1.45$ & $0.95 \pm 0.09$ \\
   Blue BAL & 4 & $38735 \pm 17$  & $+8.84 \pm 3.62$ & $991 \pm 6$ & $6.26 \pm 0.03$ & -\\
   & 5 & $38475 \pm 90$ & $-0.88 \pm 0.31$  & $865 \pm 93$ & $3.22 \pm 0.23$ & - \\
  & 6 & $38769 \pm 63$ & $+2.95 \pm 1.10$  & $801 \pm 63$ & $2.59 \pm 0.14$ & -\\
   & 7 & $38406 \pm 52$ & $-3.42 \pm 0.76$  & $994 \pm 52$ & $2.99 \pm 0.11$ & - \\
   \hline
 \end{tabular}
 \begin{tablenotes}\footnotesize
    \item[a] Spectroscopic observations in chronological order in time.
    \item[b] Centroid of the gaussian fits to the BAL components as explained in section~\ref{sec:red_bal}.
    \item[c] One sigma width of the gaussian fits to the BAL components.
    \item[d] Equivalent width of \civ \ as calculated from the gaussian fits.
    \item[e] Equivalent width of Si~{\sc iv} as calculated from the gaussian fits.
 \end{tablenotes}
\label{tab_gauss}
\end{threeparttable}
\end{table*}

\begin{table}
 \caption{Results of cross-correlation analysis of the "blue" BAL }
\centering
 \begin{tabular}{cccc}
  \hline
  Epochs& velocity& acceleration \\
  & shift (\kms) &  (cm s$^{-2}$) \\
  \hline\hline
 2,3& $654^{+194}_{-213}$ & $2.81^{+0.83}_{-0.91}$\\
 2,4& $837^{+27}_{-43}$ & $2.61^{+0.08}_{-0.13}$ \\
 2,5& $639^{+95}_{-106}$ & $1.04^{+0.15}_{-0.17}$ \\
 2,6& $974^{+81}_{-80}$ &  $1.36^{+0.11}_{-0.11}$ \\
 2,7& $530^{+33}_{-78}$ & $0.64^{+0.04}_{-0.09}$ \\  
 \hline
 \end{tabular}
\label{tab_ccor}
\end{table}



\subsection{Is there a correlated variability?}
\label{sec:corvar}
\begin{figure}
\centering
\includegraphics[viewport=8 30 800 620,width=0.5\textwidth,clip=true]{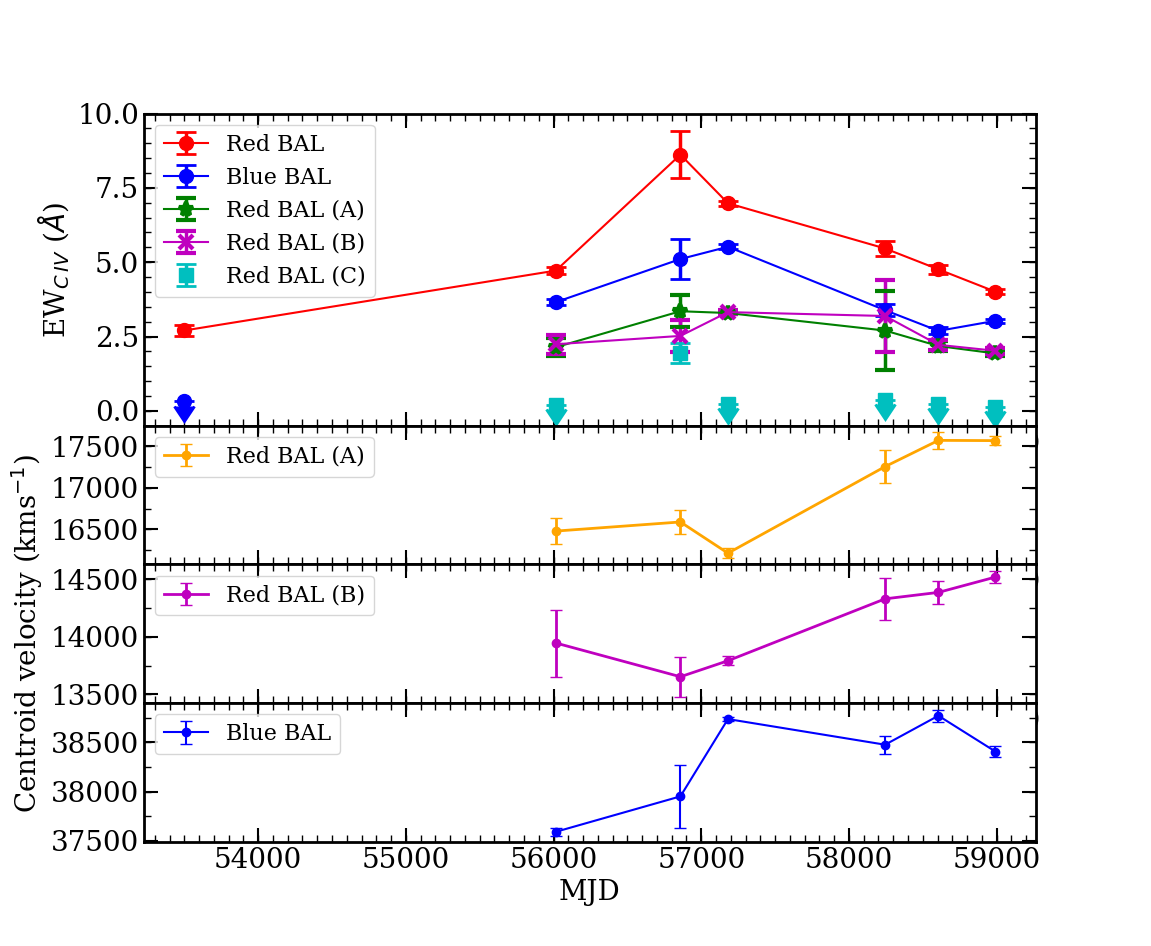}
%
\caption{{\it Top panel}: \civ\ rest equivalent widths of different absorption components as a function of time. In the case of red-BAL we plot the total equivalent width measured from the whole absorption as well as equivalent widths of individual Gaussian components (A, B and C) given in Table~\ref{tab_gauss}. The downward arrows are 3$\sigma$ upper limits when the absorption line is not clearly detected. 
{\it Bottom three panels}: Velocity centroid (3rd column in Table~\ref{tab_gauss}) of different \civ\ absorption components as a function of time.
}
\label{fig:wvar}
\end{figure}

In this section, we investigate the possible correlation between
the variations 
(both in equivalent width and kinematic shift) of the blue- and red-BAL \civ\ components. In the top panel of Fig~\ref{fig:wvar}, we plot the \civ\ equivalent widths (total as well as measured in individual Gaussian components A and B) as a function of time. The measured rest equivalent widths of \civ\ and Si~{\sc iv} at different epochs are also summarized in Table~\ref{tab_gauss}. 
It is clear from this figure that the total \civ\ equivalent width of the red-BAL increases during the first 3 epochs and then shows a steady decline afterwards. 
%
Interestingly, individual \civ\ equivalent widths of components A and B of the red-BAL 
show similar trend. 
As can be seen from Table~\ref{tab_gauss} and Fig.~\ref{fig:wvar}, the \civ\ equivalent widths of A and B are consistent with one another within errors. However, the overall profile shape does not remain constant during various epochs (see Fig~\ref{fig:absvar}). Both these components reach their maximum \civ\ equivalent width during epoch-4.  
It is also evident from the top panel of Fig.~\ref{fig:wvar} that the \civ\ equivalent width of the blue-BAL follows the trend of the red-BAL. Thus there appears to be a correlation between the blue- and red-BAL in the time evolution of the \civ\ equivalent width.

In the bottom panels of Fig.~\ref{fig:wvar}, we plot the ejection velocity of the red-BAL components A and B as a function of time. It is evident that both components A and B have blue-shifted between epoch 4 and 5. However, the amount of shift is not the same for both components (i.e 1045$\pm$208 \kms and 537$\pm$189 \kms for A and B respectively based on the gaussian centroids). This has led to significant increase in the velocity separation between the two components. This also implies that the acceleration (i.e either velocity change or directional change) is not the same for both the components. After epoch-5 there is no significant change in the ejection velocity of the components A and B while their equivalent widths have shown a steadily declining trend. 
 It is also clear from Fig~\ref{fig:wvar} (and the discussions presented in the previous section) that blue-BAL has gone through an accelerated phase between epoch-2 and epoch-4. 
However, the maximum velocity for the blue-BAL component is achieved much earlier than that for red-BAL components. The maximum delay for red-BAL to reach its maximum velocity is  
$\sim 338$ days in the quasar rest frame.

It seems there is some correlation between the variability of
the red and blue BALs.
If ionization changes are involved in these variabilities we should observe some variability of the emission line flux. We explore this in the following section.

\subsection{Broad emission line variability}
\label{sec:bel_var}

\begin{figure*}
    \centering
    \includegraphics[viewport=100 30 1200 600,width=1.0\textwidth,clip=true]{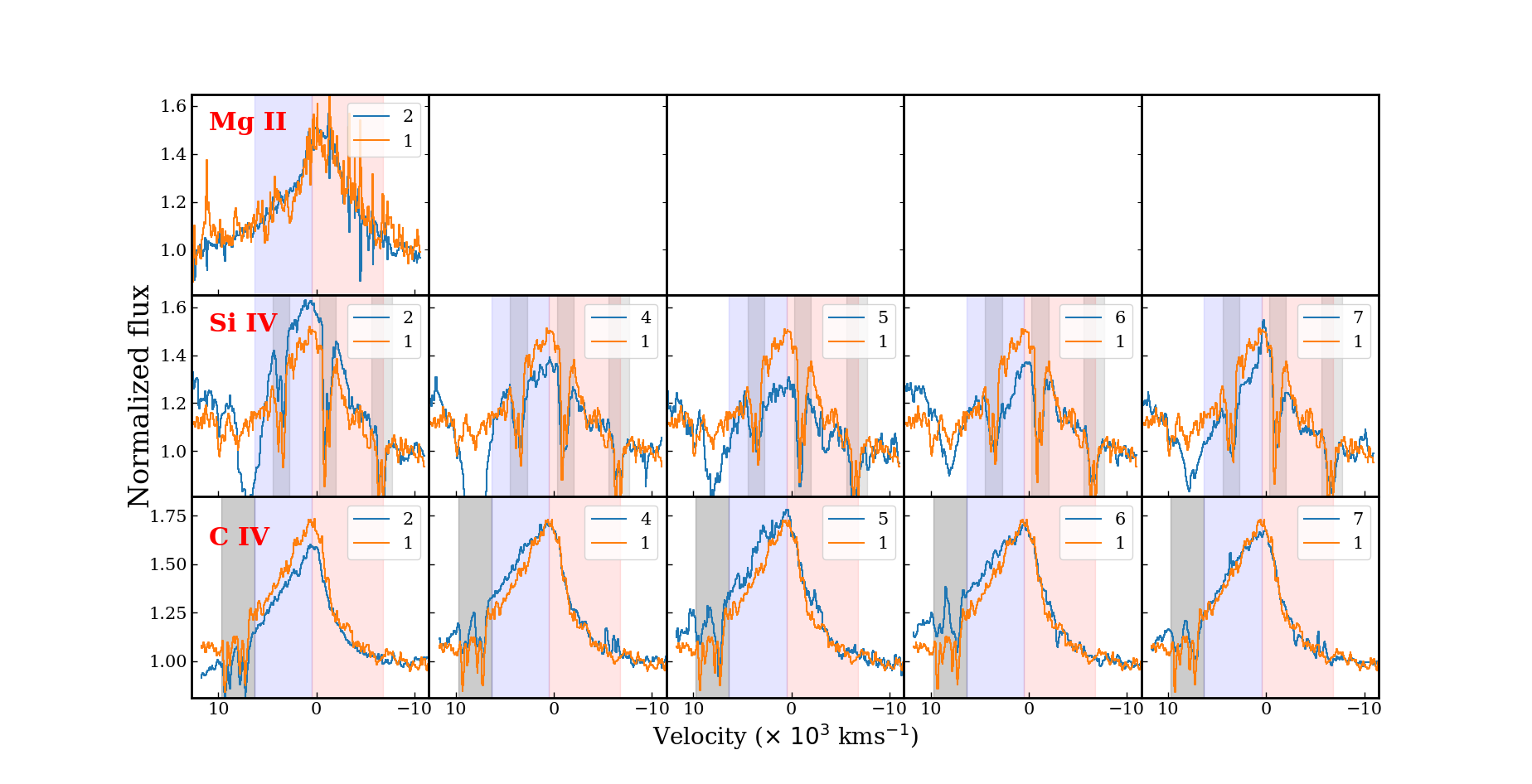}
    \caption{Comparison of emission profiles of \civ\ (bottom), \siv\ (middle) and \mgii\ (top) observed at different epochs with the corresponding profiles obtained during epoch-1. All spectra are normalised to the median flux measured over a common rest frame wavelength range.
    %
    Spectral ranges affected by 
    intervening absorption lines are marked with gray shaded regions. Red and blue part of the \civ\ emission lines are indicated with red and blue shaded regions. The figure illustrates the blue asymmetry of the \civ\ emission, the time variability of the profile and the fact that \civ\ and \siv\ equivalent widths vary in the opposite direction. The velocity scale is defined with respect to the systemic redshift (i.e $z = 2.13945$).
    }
    \label{civ_bel_var}

\end{figure*}


\begin{figure}
    \centering
    \includegraphics[viewport=50 10 950 600,width=0.5\textwidth,clip=true]{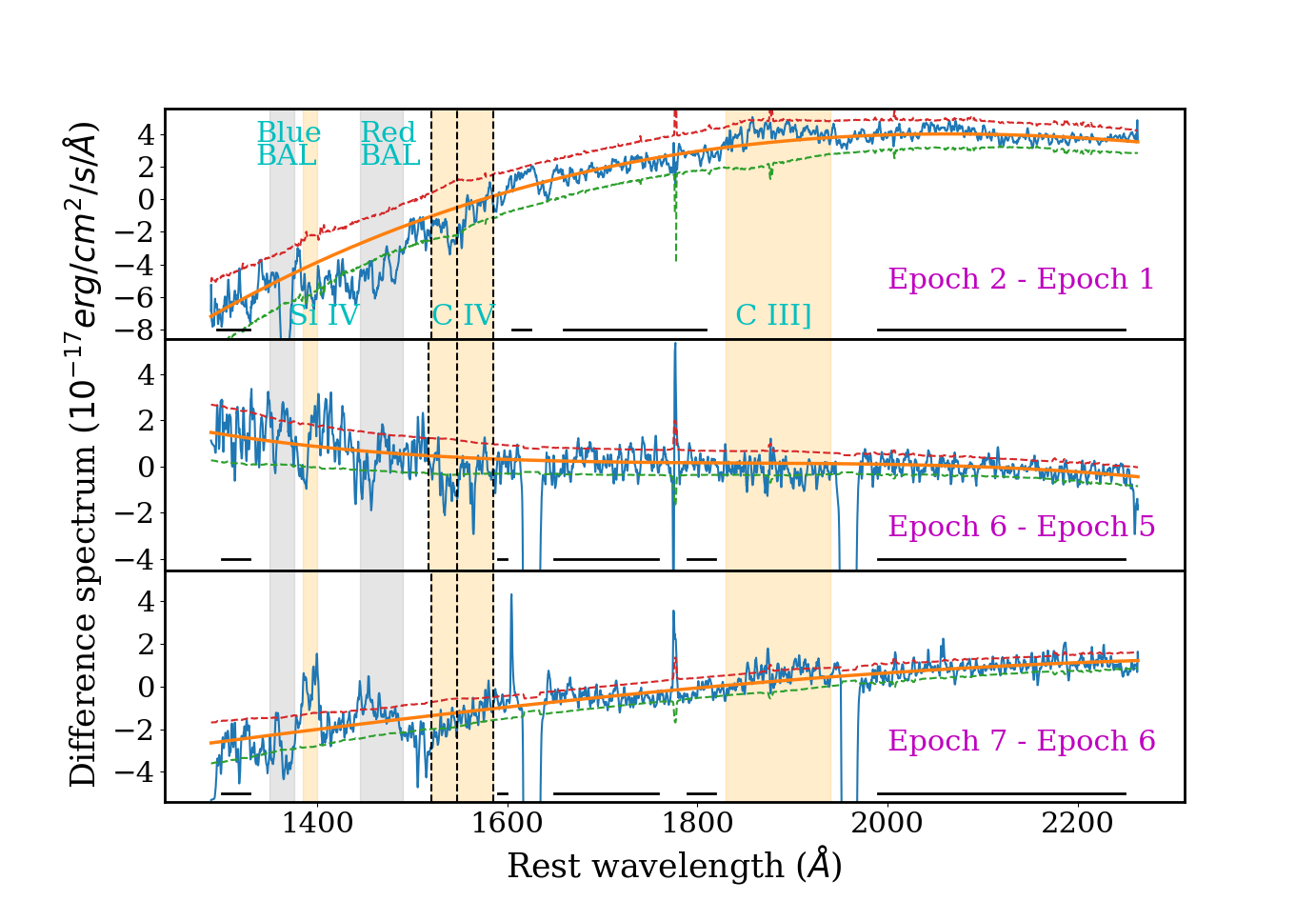}
    \caption{
Difference of the \J16 spectra observed at epoch-1 and 2 (top),  epoch-5 and 6 (middle) and epoch-6 and 7 (bottom).
The smooth 3rd order polynomial fit and associated 1$\sigma$ flux errors are shown as solid and dotted curves respectively. The regions used for the polynomial fit are shown by the horizontal black lines. The yellow and gray shaded regions show the wavelength range affected by emission and absorption lines.
}
    \label{fig:dif_spec}
\end{figure}

%
%


In this section, we study the variability of the broad emission lines.
As we do not have accurate flux calibration (due to unknown slit losses) for our SALT spectrum first we look at the 
\civ\ line equivalent width variations. For this,
we normalize all the spectra (excluding the one obtained with NTT) around the \civ\ emission line using the continuum flux measured over the rest wavelength range 1585-1615 \AA.
%
%
The normalized \civ\ emission profile for different epochs are compared with that from epoch-1 in the bottom panel of Fig.~\ref{civ_bel_var}.
Similar plots for \siv\ and \mgii\ are shown in the middle and top panels of this figure respectively. 
The wavelength range of \siv\ emission is severely contaminated by intervening absorption and the  \civ\ absorption from the blue-BAL.
Therefore, the normalization is not perfect in the case of the Si~{\sc iv} line.
The \mgii\ emission line region is covered only in the SDSS spectra. 
 In Table~\ref{BEL_table}, we summarize the emission line equivalent widths measured for the blue (i.e +500 to +6350 \kms with respect to the systemic redshift) and red (i.e -6800 to +500 \kms with respect to the systemic redshift) side of the \civ\ emission peak. The chosen regions are of unequal width because we avoid the spectral range contaminated by intervening absorption (shaded regions in Fig.~\ref{civ_bel_var}). 

It is apparent from Fig.~\ref{civ_bel_var} and Table~\ref{BEL_table} that the  \civ\ emission profile is asymmetric and skewed (i.e more flux) towards the blue (i.e low wavelength) side. 
%
Such asymmetric profiles are usually attributed to the presence of disk-winds in the BLR \citep[see for example,][]{Richards2011}.   This is the main reason for us to consider the red and blue part of \civ\ emission  separately.
%
The \mgii\ profile is contaminated by sky subtraction residuals. However we do see a slight asymmetry of the \mgii\ profile in Fig~\ref{civ_bel_var}. Moreover, when we use the same velocity range as that of \civ\ we do see the blue side having less flux compared to the red side. 
%
%

From Fig~\ref{civ_bel_var},  we can see an initial decline, then an increase followed by a decreasing trend for the \civ\ equivalent width. The Si~{\sc iv} equivalent width seems to follow the opposite trend.
It is also evident from Table~\ref{BEL_table} that both red and blue part of the \civ\ equivalent width show similar trend with variation in the blue part being larger. This is also reflected in the fact that the equivalent width ratio of blue and red parts (last column in Table~\ref{BEL_table}) follows the trend with time seen in the \civ\ rest equivalent width.
From the top panel in Fig~\ref{civ_bel_var}, we can see no clear variation for the Mg~{\sc ii} profile between the first two SDSS epochs. 

The emission line equivalent width variations can be driven by line and/or continuum variations. 
To quantify the flux variations, we first consider the spectra from first two-epochs (i.e SDSS spectra). In the top panel of Fig~\ref{fig:dif_spec} we show the difference between the two spectra as a function of wavelength. While calculating the difference spectrum we applied the correction to the BOSS spectrum as suggested by 
\citet[][using their equation 1]{Guo2016}. In this plot we also mark the regions affected by emission and absorption. It is evident that the continuum shape has changed with more variations in the blue. We fit this difference spectrum with a 3rd order polynomial. 
By integrating the residual spectrum (after subtracting the fit) over the shaded region marked as \civ\ in Fig.~\ref{fig:dif_spec},
we estimate the reduction in the \civ\ flux to be (4.82$\pm$0.69)$\times10^{-16}$ erg s$^{-1}$ cm$^{-2}$ between epochs 1 and 2. This corresponds to a \civ\ line flux variation by $\sim$11\% between epochs 1 and 2.
We also find that the \siv\ line flux has reduced by $(1.59 \pm 0.31)$ $\times10^{-16}$ erg s$^{-1}$ cm$^{-2}$  between epoch 1 and 2.
Therefore, the increase in \siv\ equivalent width we note in Table~\ref{BEL_table} is dominated by the change in the continuum flux around the \siv\ emission.

We have access to g- and r-band photometric measurements on the nights of last three SALT observations (see Fig.~\ref{fig:ztfslc}).
Using this we estimate the slit loss and obtain the absolute flux scale for these epochs. The g- and r-magnitudes have error of $\sim$0.03 mag. 
%
The difference of the spectra obtained during the last two SALT epochs (6 and 7)  after correcting flux levels is shown in the bottom panel of Fig.~\ref{fig:dif_spec}.
The \civ\ line flux has decreased between epoch 6 and 7 by (1.59$\pm$0.28)$\times10^{-16}$ erg s$^{-1}$ cm$^{-2}$. This corresponds to a \civ\ line flux variation by 4\%. Given the flux uncertainties this may not be a significant change.
In the same way, we find that the \civ\ flux has been reduced between epoch-5 and 6 by (5.03$\pm$0.30)$\times10^{-16}$ erg s$^{-1}$ cm$^{-2}$  (see the middle panel in Fig.~\ref{fig:dif_spec}). This is $\sim$13\% change in the \civ\ line flux. Thus during the last three spectroscopic epochs (where we have photometry to get absolute flux scales) we see the \civ\ emission line flux continuously decreasing by $\sim$17\%. Unfortunately we do not have photometric measurements during epoch-4 when the \civ\ equivalent width was maximum. 

In low-$z$ AGNs \civ\ is known to show a positive response to the continuum variations with some time-delay. Also \civ\ and Si~{\sc iv} tend to show similar responses \citep[see ][]{derosa15}. 
In these AGNs typical continuum variations are of the order of
20\% when emission lines variations are less.
%
Our finding would imply large ionizing flux variations, as the BLR size of \J16 is expected to be much larger than that of low-z low luminosity AGNs. 
Such variations in the ionizing flux could also help us understand the origin of the absorption line variability.

\begin{table}
\begin{threeparttable}
\caption{Broad emission line variability}
 \begin{tabular}{ccccc}
  \hline
   Ion & Epoch & EW (blue) \tnote{a} & EW (red) \tnote{b} & $\frac{EW (blue)}{EW (red)}$ \\
    & & (\AA) & (\AA) & \\
   \hline
   \hline
   & 1 & 13.72 $\pm$ 0.19 & 8.83 $\pm$ 0.21 & 1.55 $\pm$ 0.04\\
   & 2 & 10.86 $\pm$ 0.13 & 7.19 $\pm$ 0.14 & 1.51 $\pm$ 0.03 \\
   & 4 & 15.71 $\pm$ 0.10 & 9.69 $\pm$ 0.10 & 1.62 $\pm$ 0.02\\
   \civ\  & 5 & 16.62 $\pm$ 0.22 & 10.07 $\pm$ 0.22 & 1.65 $\pm$ 0.04 \\
   & 6 & 15.64 $\pm$ 0.14 & 9.36 $\pm$ 0.14 & 1.67 $\pm$ 0.03\\
   & 7 & 14.12 $\pm$ 0.08 & 8.82 $\pm$ 0.09 & 1.60 $\pm$ 0.02 \\
   \hline
   & 1 & 4.09 $\pm$ 0.13 & 2.43 $\pm$ 0.15 & 1.68 $\pm$ 0.12 \\
   & 2 & 5.49 $\pm$ 0.10 & 3.73 $\pm$ 0.11 & 1.47 $\pm$ 0.05\\
   & 4 & 3.04 $\pm$ 0.06 & 2.19 $\pm$ 0.08 & 1.39 $\pm$ 0.07\\
   \siv\ & 5 & 2.27 $\pm$ 0.14 & 1.16 $\pm$ 0.18 & 1.96 $\pm$ 0.33\\
   & 6 & 2.40 $\pm$ 0.09 & 2.36 $\pm$ 0.11 & 1.02 $\pm$ 0.06\\
   & 7 & 3.11 $\pm$ 0.06 & 1.94 $\pm$ 0.07 & 1.60 $\pm$ 0.10\\
   \hline
   \mgii\ & 1 & 11.00 $\pm$ 0.63 & 20.38 $\pm$ 0.92 & 0.54 $\pm$ 0.04\\
   & 2 & 11.99 $\pm$ 0.32 & 18.06 $\pm$ 0.46 & 0.66 $\pm$ 0.02 \\
   \hline
 \end{tabular}   
 \begin{tablenotes}\footnotesize
    \item[a] Equivalent width calculated for the velocity range 500 to 6350 kms$^{-1}$ blueward of the systemic redshift for \civ\ and \mgii\ and 500 to 2600 kms$^{-1}$
    for \siv.
    \item[b] Equivalent width calculated for the velocity range 500 to -6800 kms$^{-1}$ redward of the systemic redshift for \civ\ and \mgii\ and -2100 to -5500 kms$^{-1}$ for \siv.
 \end{tablenotes}  

\label{BEL_table}
\end{threeparttable}
\end{table}

\section{Continuum variability:}
\label{sec:continuum}

\begin{figure*}
\centering
\includegraphics[viewport=70 30 950 600,scale=0.57,clip= true]{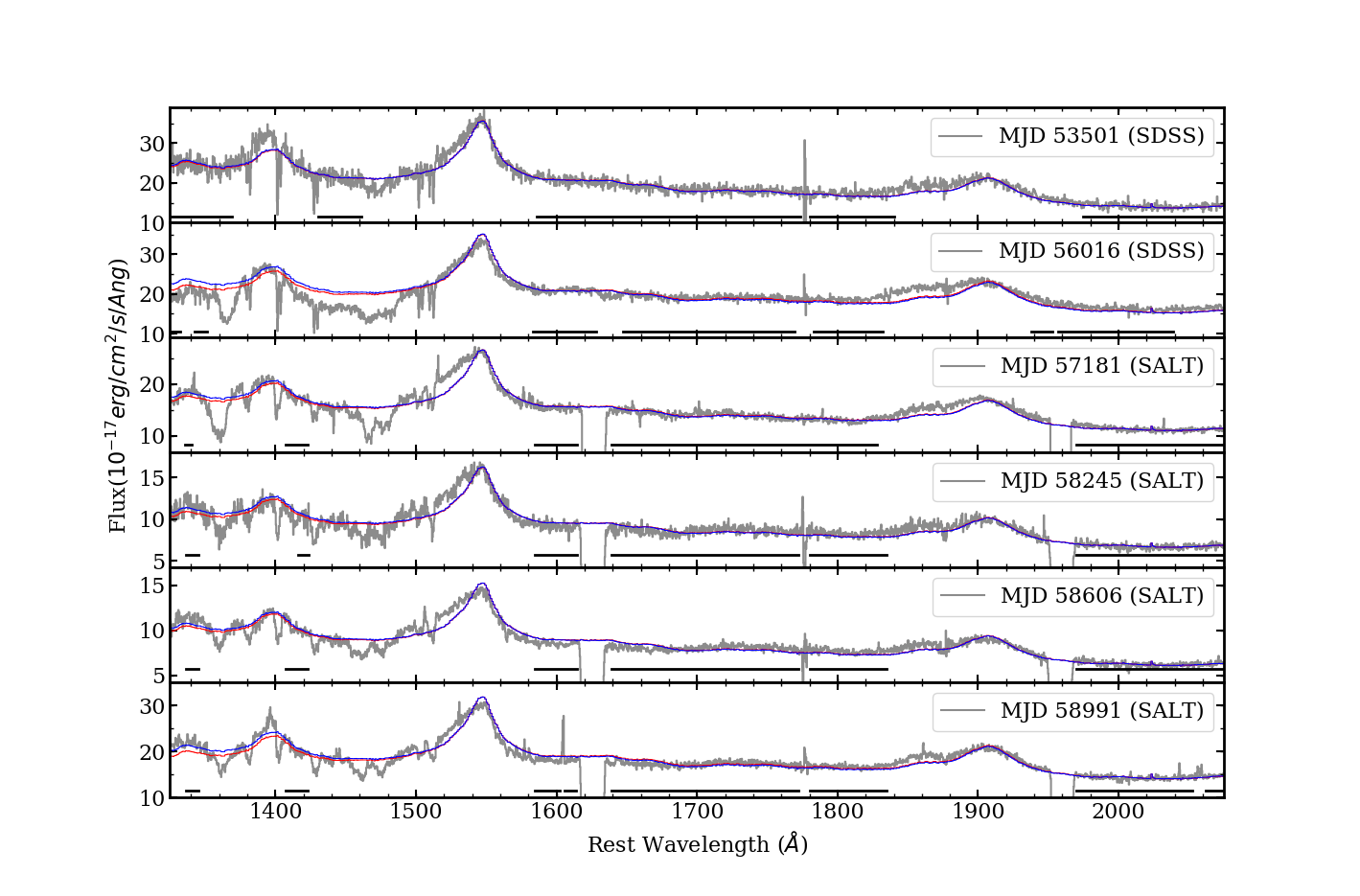}
\caption{Observed spectra of \J16 at different epochs. The spectrum obtained with 
NTT at epoch-3 is not shown as it was not possible to flux calibrate that spectrum.
Wavelengths are given in the rest frame of the quasar at \zem = 2.13945. 
The red and blue curves are the results of the fit to the observations by
a quasar template attenuated by dust or tilted 
by a power-law, respectively,
as discussed in Section~\ref{sec:continuum}. As expected, the two corrections are degenerated.
The black horizontal lines mark the regions devoid of any absorption or emission lines which are used during the continuum fitting. Spectral gaps  seen
in the SALT spectra around $\lambda\sim1630$~\AA\ and 1960~\AA\ are due to physical gaps between the three CCDs of RSS.
}
\label{fig:continuum}
\end{figure*}

In this section, we focus on the continuum (i.e flux as well as color) variability of \J16. For this purpose we will use continuum fits to the observed spectra as well as photometric light curves.

%

\subsection{Optical continuum and color variations}
\label{sec:photo}


%
%
%

We quantify the variability in the spectral shape by fitting the observed flux-calibrated spectra with a quasar template
obtained from the SDSS spectra of 186 non-BAL quasars 
at $2.0 < $\zem$ < 2.25$ having r- and i-band magnitudes within 0.2 mag of \J16\ magnitudes measured during the SDSS photometric survey. 
Any difference found between individual spectra and the template will be ascribed to
dust-extinction or to the particular slope of the quasar continuum.


We reddened the composite spectrum using different (LMC, SMC and Milkyway) extinction curves given in \citet[][]{Gordon2003} and the method explained in \citet{srianand2008}. We consider three possible locations for the dust, at \zem\ and at the redshifts of the ``red-" and ``blue-BAL" components mentioned above.
We use a $\chi^2$-minimization code to obtain the best fitted values of two parameters, a constant normalization to match the observed flux and dust extinction in the V-band ($A_{\rm V}$) to match the spectral shape. For this fit, we consider only those regions devoid of emission or absorption lines in the spectra (see Fig.~\ref{fig:continuum}). 
We find 
that in all cases the fit obtained using the SMC-like extinction curve has the minimum $\chi^2$ irrespective of the dust location. Also for a given extinction curve, dust located at \zem\ is favored.
The best fitted values of $A_{\rm V}$ are summarized in column 9 of Table~\ref{tab_obs}. The main error in $A_{\rm V}$ comes from the error in $R_{\rm V}$. For the $1\sigma$ range in $R_{\rm V}$ found for SMC \citet[i.e between 2.61 to 2.87, as given in ][]{Gordon2003}
the typical error in $A_{\rm V}$ is of the order of 0.01. 
From Table~\ref{tab_obs}, it is clear that if the spectral shape is governed by dust extinction then $A_{\rm V}$ is changing between different epochs. In particular, we find a factor of $\sim$5 increase in $A_{\rm V}$ between two SDSS epochs (i.e epoch-1 and epoch-2) during which time we also notice that the quasar shows a {\it redder-when-brighter} trend.

%

While we have fitted the continuum using variations in $A_{\rm V}$ it is possible that changes in the power-law spectral index also cause such a variation.
To mimic this, we multiply the composite spectrum by a power-law of the form $C (\frac{\lambda}{\lambda_0})^p$. Here $C$, $\lambda_0$ and $p$ are free parameters. We find the best fit parameters ($C$ and $p$) using  $\chi^2$-minimization. 
The best fit values of $p$ for different epochs are summarised in column ten of Table~\ref{tab_obs} and the fits are shown in Fig~\ref{fig:continuum}. 
This confirms that the spectral shape changes significantly
between different epochs.
As expected there is a tight correlation between $A_{\rm V}$ and $p$.




\subsection{Light curves and photometric variability}
\label{sec:lc}

\begin{figure}
    \centering
    \includegraphics[viewport=30 30 1000 600,scale=0.29,clip=true]{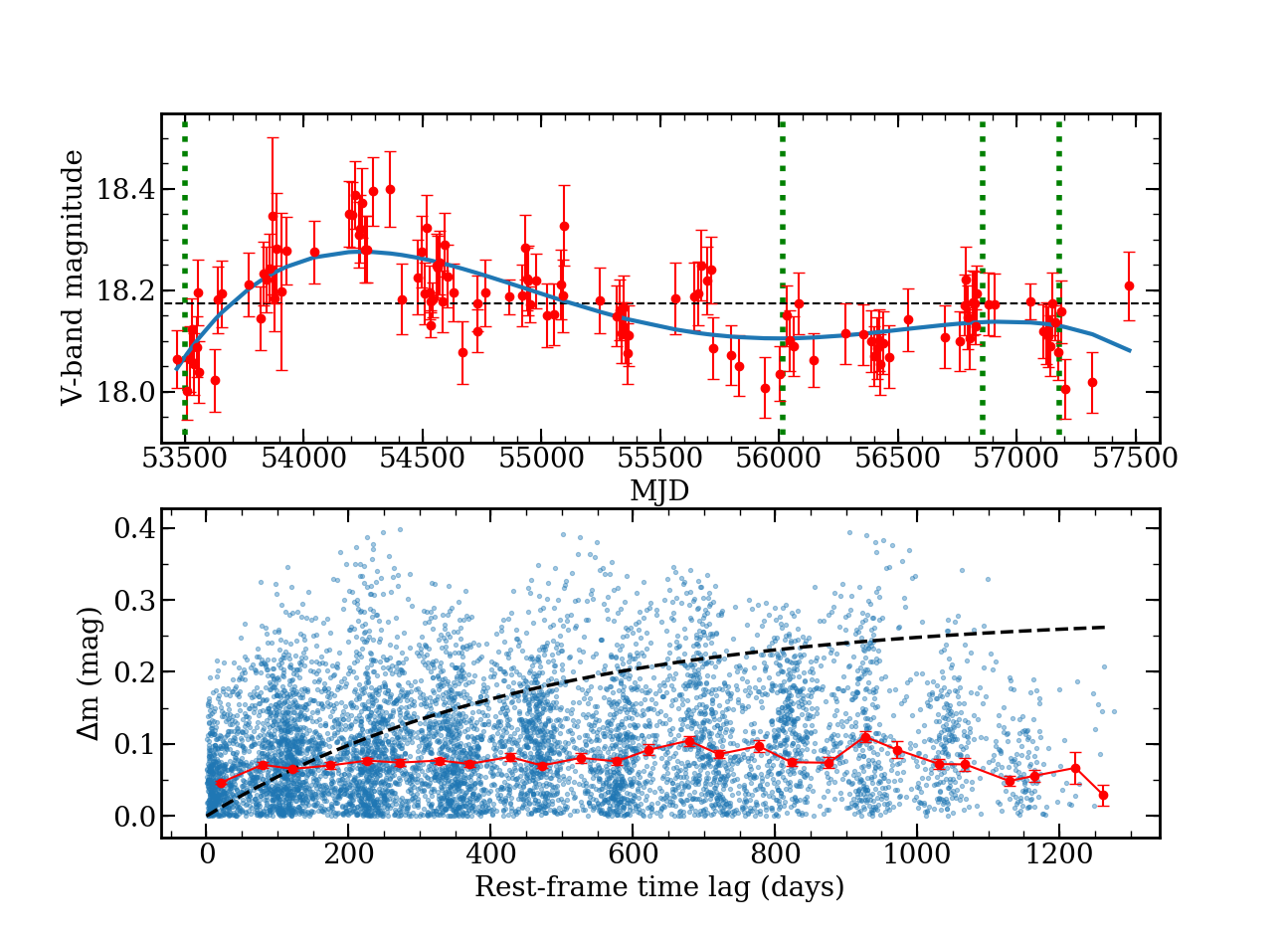}
    \caption{{\it Top panel:} V-band light curve of \J16 from CRTS with a low order polynomial fit overlaid. The vertical dotted lines indicate the dates of the first four spectroscopic epochs. The horizontal line gives the median V-band magnitude of \J16 during this period. {\rm Bottom panel:} Structure function obtained from the light curve shown in the top panel. The red points are the average values in a rest frame time-lag bin of 50 days. 
    The best-fit to the structure function for SDSS quasars is shown as a black dashed line \citep{MacLeod2012}. The flat structure function over 1200 days suggests that there is no fading or brightening of \J16 over these timescales.}
    \label{fig:crtslc}
\end{figure}

\begin{figure}
    \centering
    \includegraphics[viewport=10 30 1000 670,scale=0.32,clip=true]{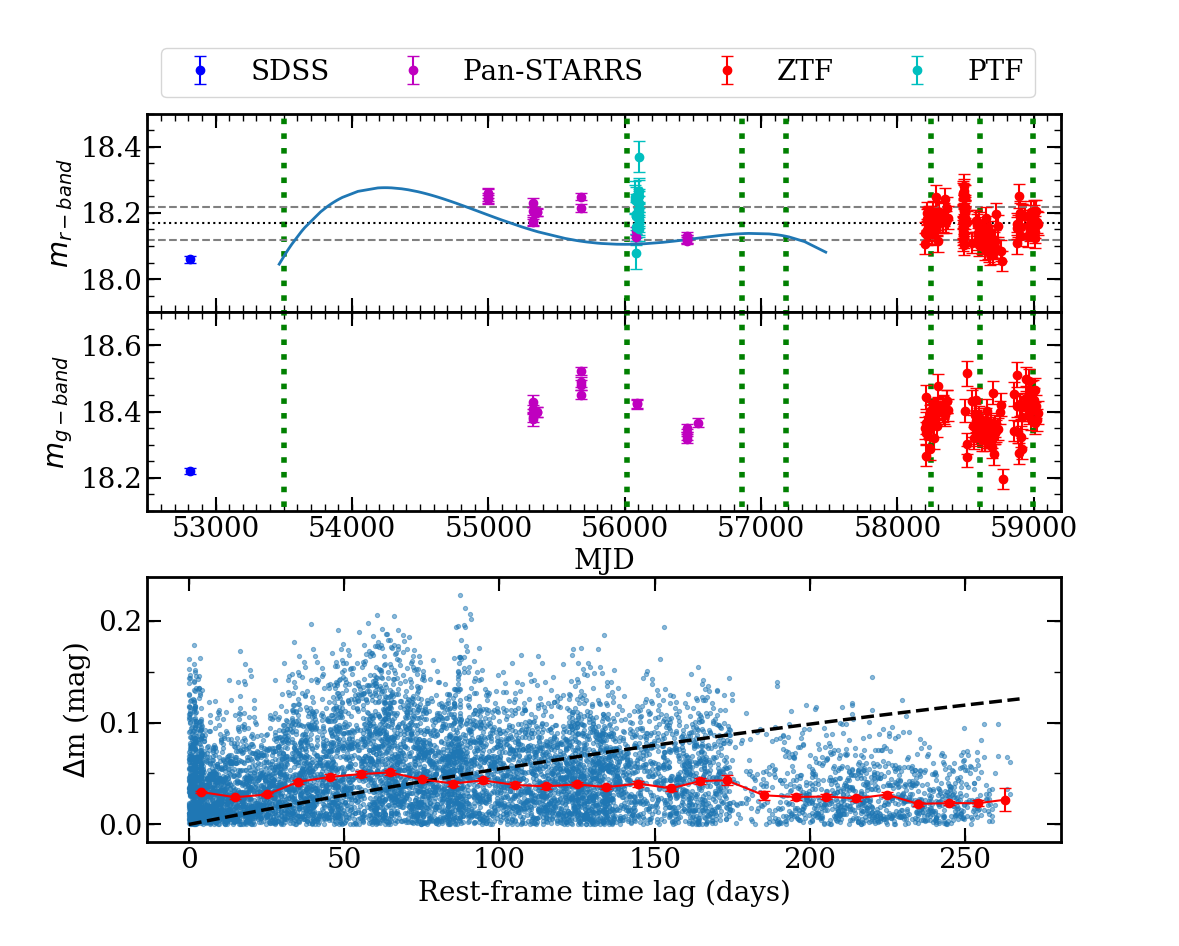}
    \caption{{\it Top panel:} r-band magnitudes of \J16 obtained from SDSS, Pan-STARRS, ZTF and PTF. The curve is the fit to the V-band light curve from CRTS as shown in Fig.~\ref{fig:crtslc}. The dotted and dashed horizontal lines indicate a r-band magnitude of 18.17$\pm$0.05 mag.
    {\it Middle panel:} g-band magnitudes of \J16 obtained from SDSS, Pan-STARRS and ZTF.
    {\it Bottom panel:} the structure function obtained from the r-band light curve from ZTF. There is no indication of coherent brightening or fading of \J16 over  time-scales of 100 to 250 days. 
    }
    \label{fig:ztfslc}
\end{figure}

We also collected publicly available photometric light-curves of \J16 from
the Panoramic Survey Telescope and Rapid Response System \citep[Pan-STARRS;][]{panstarrs2016}, The Palomar Transient Factory \citep[PTF;][]{Law2009} and the Zwicky Transient Facility \citep[ZTF;][]{zptf2019b,zptf2019a} surveys. PTF provides phototometric data only in the r band. Pan-STARRS provides photometric data of \J16 for five broad band filters, i.e, g, r, i, z and y whereas ZTF gives the same for the g, r and i bands. In both these cases the observations in different bands were not obtained simultaneously and are typically separated by up to a few days.  Photometric observations from the SDSS have also been considered. 

Catalina Real-time Transient Survey \citep[CRTS,][]{drake2009} provides the light curve covering nearly 11 yrs (i.e $\sim 3.5$ years in the rest frame of \J16, covering our first four spectroscopic epochs). CRTS operates without any specific filter and the resulting open magnitudes are converted to V band (corresponding to a rest wavelength of $\lambda \simeq 1750$~\AA\ in the quasar frame) magnitudes using the transformation equation $V=V_{\rm ins}+a(v)+b(v)\times(B-V)$, where $V_{\rm ins}$ is the observed open magnitude, $a(v)$ and $b(v)$ are the zero-point and the slope. These are obtained from three or more comparison stars in the same field with the zero-point error typically being of the order of 0.08 mag. CRTS provides four such observations 
taken 10 min apart on a given night. Since we are mainly interested in the long-term variability, we take the median magnitude of these four points (or less if less points are available)
to get the light curves shown in Fig~\ref{fig:crtslc}. 

Note that CRTS light curves, while sensitive to overall variations in the quasar brightness, will not capture colour variations accurately. Also these light curves will be influenced by any emission and absorption line variations. It is interesting to note that the inferred V-band magnitudes close to the two SDSS spectroscopic epochs (i.e epoch-1 and epoch-2) are nearly identical. However, the quasar has faded by 0.3 mag (within a year in the quasar's rest frame after epoch-1) before brightening back to the first SDSS epoch  value by the time epoch-2 spectrum was observed by SDSS. It is interesting to note, from the smooth polynomial fit (continuous curve in Fig~\ref{fig:crtslc}) to this light curve, that  the V-band magnitudes are nearly same during the subsequent two spectroscopic epochs (i.e epoch-3 and epoch-4).
Just to quantify the small time-scale (i.e $<1$ year in the quasar's frame) variability we construct the structure function for the CRTS V-band magnitudes using the method described in \citet{MacLeod2012}. For rest frame lag time-scale probed (i.e up to 1200 days)  the structure function is nearly flat at 0.05 mag. We do see a slight excess around a lag time of $\sim$900 days. This basically suggests that there is no systematic brightening or fading  (i.e by more than say 0.1 mag) of \J16\  over the  time-scales we are probing in our spectroscopic monitoring.  In the bottom panel of Fig.~\ref{fig:crtslc}, we also plot the best fitting curve for the structure function of SDSS quasars obtained by \citet{MacLeod2012}. It is clear that \J16\ vary much less than typical SDSS quasars over a rest frame time-scale of $200-1200$ days.


 The g- and  r-band (corresponding to a rest wavelength of $\lambda\simeq 1530\pm220$\AA\ and $\lambda\simeq1990\pm220$\AA\ respectively) light curves of \J16 constructed using photometric observations from Pan-STARRS, PTF, ZTF and SDSS are shown in the top two panels of Fig~\ref{fig:ztfslc}. Unlike the light curve obtained with CRTS, this combined r-band light curve sparsely samples the spell of time of 
the first 4 spectroscopic epochs. ZTF provides much better time sampling over the last three spectroscopic epochs.
The dotted and dashed horizontal lines show a range in r-magnitude of  $r=18.17\pm 0.05$ mag. Most of the observed points are within this range. This again confirms the lack of large coherent r-band magnitude variations over large time-scales apart from the $\sim$0.3 mag dimming seen in the CRTS light curve (see Fig~\ref{fig:crtslc} and the curve shown in Fig~\ref{fig:ztfslc}). Also the r-band magnitudes measured close in time to the spectroscopic monitoring periods (i.e vertical dotted lines) are nearly same. 
%
The structure function computed from ZTF r-band observations are shown in the bottom panel of Fig.~\ref{fig:ztfslc}. This also confirms that the long-term rest frame UV variability of \J16\ is weaker than what is found for typical SDSS quasars over a quasar rest frame time-scale of $\ge$100 days.


\subsection{Color variability:}
\label{sec:colvar}
\begin{figure}
    \centering
    \includegraphics[viewport=30 20 1000 570,scale=0.29]{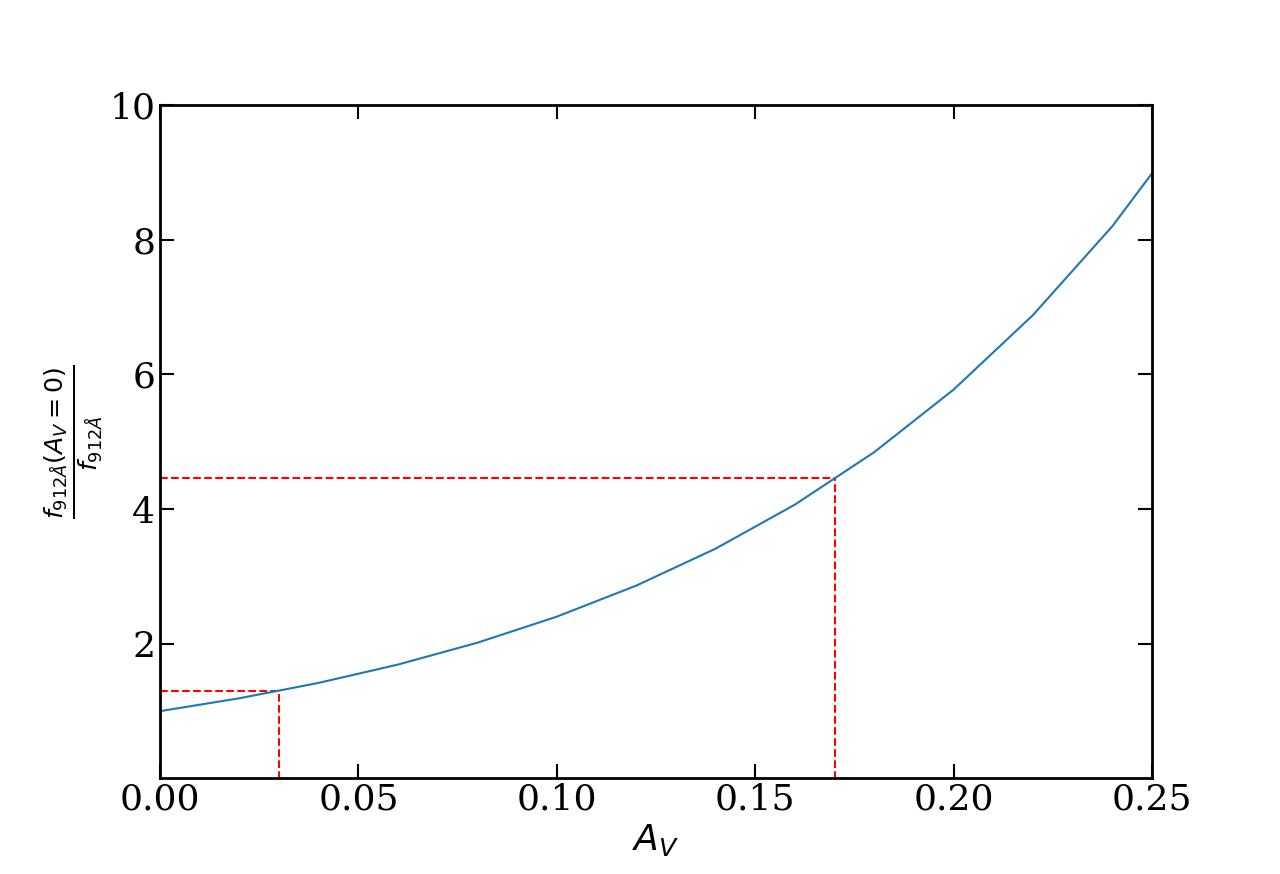}
    \caption{The ratio of the flux at $912$ \AA \ without applying any dust extinction (i.e, the dust optical depth at V-band, $A_{\rm V} = 0$) to the flux at 912$\AA$ after applying dust extinction with $A_{\rm V}$ as given 
    along the X-axis.
    The red dashed lines indicate the value of this ratio at $A_V$ = 0.03 and 0.17 mentioned in Section~\ref{sec:continuum}
    } 
    \label{f912_AV}
\end{figure}

%

In section~\ref{sec:continuum}, we used a template fitting method to derive $A_{\rm V}$ that we found to vary in the range 0.03 to 0.17 (see column 9 of Table.~\ref{tab_obs}) for the SMC extinction curve.
In Fig.~\ref{f912_AV}, we plot the ratio of the H~{\sc i} Lyman continuum flux (i.e at $\lambda = 912$\AA) with and without reddening as a function of $A_{\rm V}$. 
It can be seen that small changes in $A_{\rm V}$ can lead to large changes in the Lyman continuum flux. 
%
%
The reddening of the continuum
can lead to a factor 3.5 reduction in the hydrogen ionizing radiation between epoch 1 and 2. 
Similarly between epoch 2 and 6, purely based on changes in $A_{\rm V}$ we expect the H~{\sc i} ionizing flux to increase by up to a factor$~$2.5. 
Note the SMC extinction curve is not available for the energy range above the \civ\ ionization level. 
A simple extrapolation 
from the UV range suggests an order of magnitude change in the \civ\ ionizing flux for the above mentioned changes in $A_{\rm V}$.
%
Similarly if we assume a constant V-band flux, we expect a factor of 2 change in the 912\AA\ flux between epoch 1 and 2, due to the change in the powerlaw index. This will correspond to a factor of $\sim$4 change in the C~{\sc iv} ionizing flux. Similarly between epoch 2 and 6 we expect the 912\AA\ (respectively C~{\sc iv} ionizing) flux to increase by a factor of $\sim$1.6 (respectively 2.7). 

\begin{figure}
    \centering
    \includegraphics[viewport=40 17 1210 600,scale=0.26,clip=true]{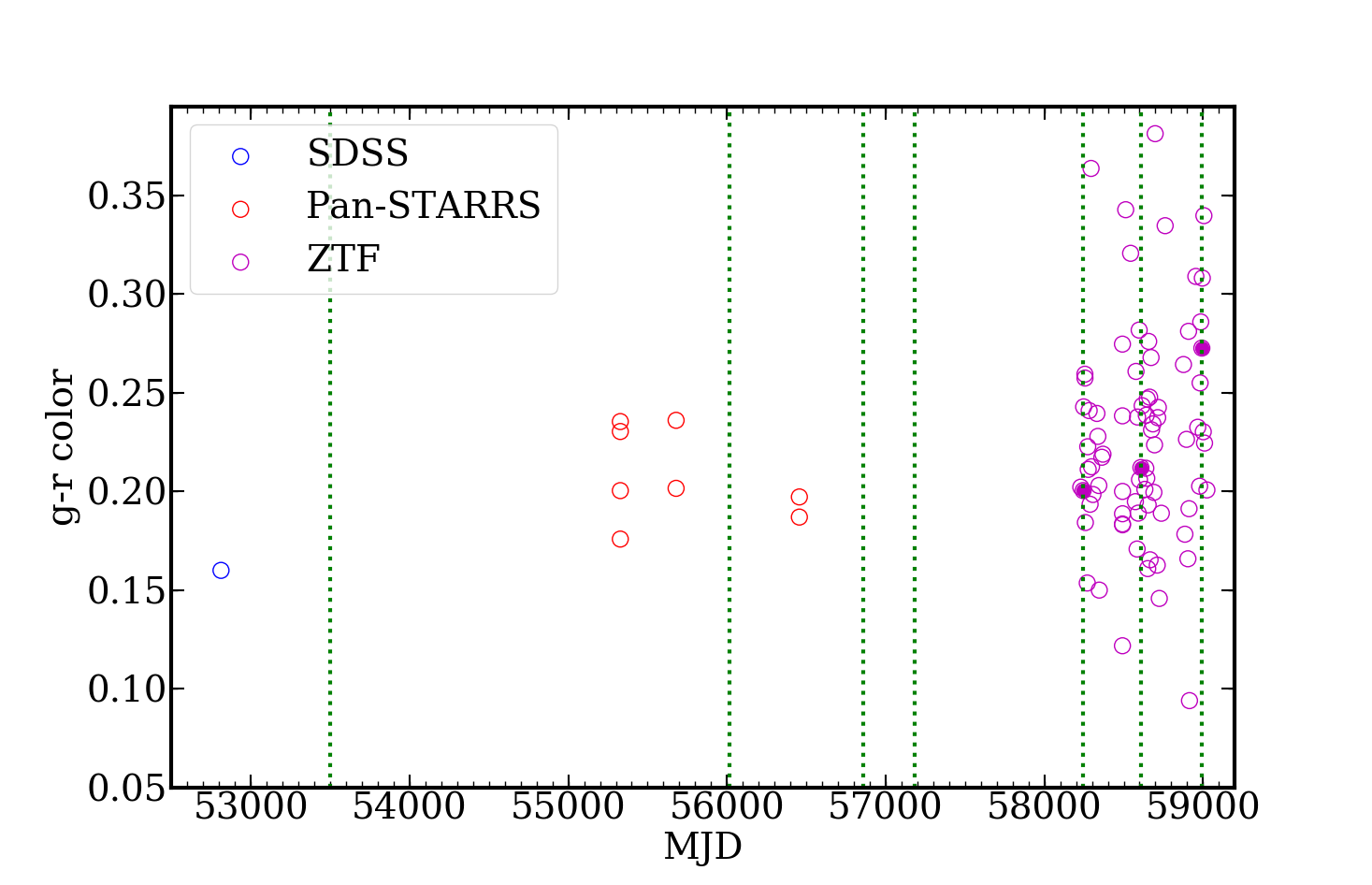}
    \caption{Plot showing g-r colour as a function of time. We consider the closest g and r measurements taken within a day for this purpose. The filled symbols are the g-r measurements obtained on the night of our spectroscopic observations (shown with vertical dotted lines). From the well measured ZTF colors (with a typical error of 0.04 mag) we do see large variations in the g-r colors within very short time-scales. }
    \label{fig:ztfcol}
\end{figure}

Confirming the color variations and their dependence on the quasar luminosity is important for quantifying the photoionization process.
Therefore, we probe the g-r color (corresponding to the spectral shape in the rest wavelength range 1310-2310~\AA) variability using available photometric observations.
Remember the g-band covers the \civ\ and Si~{\sc iv} emission lines and the \civ\ absorption related to both the blue- and red-BAL components. From our spectrum we find that the net contribution of these two emission lines to the g-band  will be $\le$6\%.
%
%
In  Fig~\ref{fig:ztfcol}, g-r color is shown as a function of time, where we considered only the nearest g and r band photometric points obtained within a day. It is evident that compared to SDSS photometric measurements the quasar appears to be redder in subsequent measurements.
It is clear from Fig.~\ref{fig:ztfcol} that there is no g-r color measurements close to the first four spectroscopic epochs. However we do have measurements (filled circles) on the nights of the last three epochs of spectroscopic observations.
We see that the g-r color varies by up to 0.3 mag (typically at $>3\sigma$ level) within a short time scale even when the r-band magnitude does not vary much (see Fig.~\ref{fig:ztfslc}).
It is also clear that  such large colour variations cannot be accounted for by the emission line variations alone (as we also see in Fig~\ref{fig:dif_spec}).

\begin{figure}
    \centering
    \includegraphics[viewport=35 12 1200 610,width=9.5cm,clip=true]{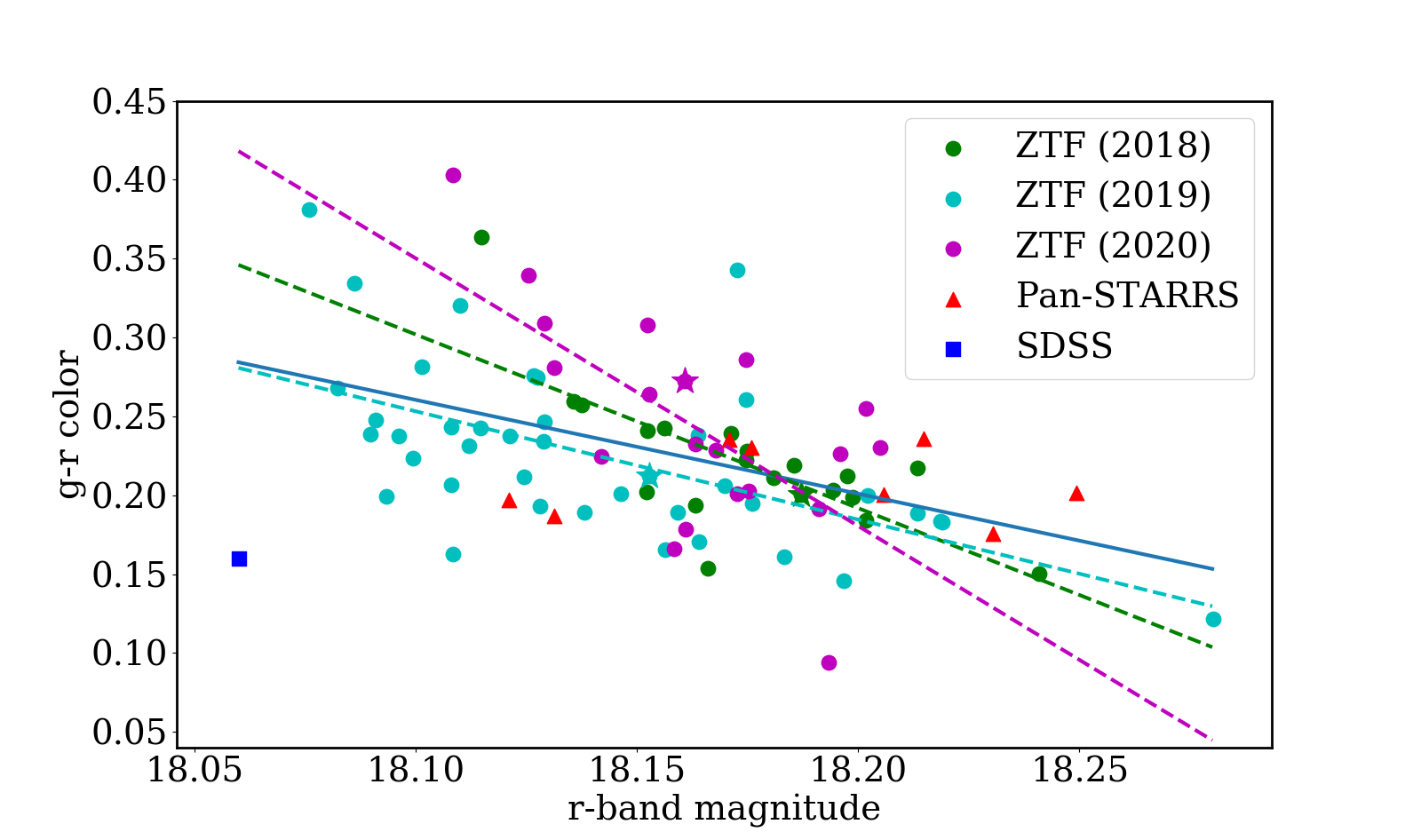}
    \caption{Plot showing the r-band magnitude vs g-r color using g- and r-band measurements within a day from ZTF, Pan-STARRS and SDSS photometric surveys. The green, cyan and magenta dashed lines correspond to linear fits of year-wise (ie, years 2018, 2019 and 2020 respectively) ZTF photometric points. The blue solid line corresponds to the linear fit to the full sample. Stars are measurements on the days when our spectroscopic observations have taken place for the last three epochs.} 
    \label{fig:rmag_color_plot}
\end{figure}

To explore the color variation further we plot r-band magnitude vs g-r colors from different observations in Fig.~\ref{fig:rmag_color_plot}. For this we consider only those g- and r-band photometric points that were obtained within a single night. Typical errors are $<0.02$ and $<0.04$ mag for the r-band magnitude and g-r colors respectively.
When more observations are available on a given night in one band we considered the nearest g-band measurement to the r-band measurement. The ZTF points obtained over the past three years dominate the plot. 
%
For the full sample (without the initial SDSS point), we notice a strong anti-correlation (with spearman rank correlation coefficient of $-0.45$) between the two quantities in the sense that when \J16 is bright in r-band it also tends to be redder.  This anti-correlation is present even when we consider data obtained each year separately. The linear regression fits to ZTF data taken during three different years are also shown by dotted lines in different colors in Fig.~\ref{fig:rmag_color_plot}. 
This anti-correlation is consistent with what we see between
the first two epoch spectra (see Fig~\ref{fig:dif_spec} and Section~\ref{sec:continuum}).
%
Moreover, the g-r colour has a large spread for a given r-band magnitude. In particular, the scatter in g-r is larger for smaller r-magnitude. This means that even if two epochs have similar r-band magnitudes their g-band magnitude (in turn the \hi\ ionizing flux) may be very different. 
%
%
%
%
%
Simultaneous (with in a night) r-band and i-band measurements from ZTF are available only for 19 epochs (from ZTF). While r-i (i.e rest wavelength range 1720 to 2600\AA) colour shows 0.2 mag
variation for a given r-band magnitude we do not find any correlation between r-i color and r magnitude. 
%
%

The ``redder-when-brighter"  (RWB) trend shown by \J16 is contrary to the general
``bluer-when-brighter" (BWB) trend shown by typical quasars during their variability \citep[][]{Wilhite2005, Schmidt2012, Ruan2014}.
Such a variability is usually attributed to global accretion rate variations or local temperature variations in a highly turbulent disk. 
Using multi-epoch spectroscopic data of 2169 quasars in SDSS DR7 and DR9, \citep{Guo2016} found that 94$\%$ of the quasars followed a BWB trend whereas only 6$\%$ followed a red-when-brighter RWB trend. They also showed that the distributions of blackhole mass, Eddington ratio, redshift and UV luminosity of sources showing RWB and BWB are statistically indistinguishable. Interestingly, in another study by \citep{Bian2012}, almost half of the sources out of a sample of 312 radio-loud (RL) quasars and 232 radio-quiet (RQ) quasars showed RWB behaviour. It was suggested that the differences between the two samples could come from jet contribution to the continuum emission, differences in the redshift distribution of sources and  time-scale probed \citep[see Section 5.5,][]{Guo2016}.
%

Unlike these studies we probe the colour variations over a wide range of time-scales. In addition our measurements 
from photometry are not affected by 
issues related to flux losses associated with variable seeing
conditions.
%
As \J16 is not detected in FIRST or NVSS we can conclude that the colour variations are not driven by the presence of 
a radio jet.
We note that in a simple model where the flux variability is caused by local fluctuations that propagate through the accretion disk, the RWB trend can be reproduced 
if the fluctuations occur first in the outer accretion disk and have not yet propagated inward to the inner accretion disk region where most of the high energy photons are created.

%
%
%
%

{\it In summary, the light curves studied here suggest that apart from one episode between epoch-1 and epoch-2 (during which \J16 experienced a continuous fading) there is no significant long term brightening or fading of the quasar over our monitoring period. 
The quasar shows color variations on short time scales
suggesting possible changes in the accretion rate and/or local thermal disturbances in the disk. The RWB trend, if
extrapolated to higher energy ranges that are not probed by
our spectra, can lead to a significant reduction in the ionizing flux when the r-band flux increases. 
}


\section{Discussions}
\label{sec:discuss}


\subsection{\civ\ broad emission line variations}

In the standard picture, emission line variations are usually attributed to changes in the ionising field.
In section~\ref{sec:bel_var}, we observe that the \civ\ emission line equivalent width varies significantly over our monitoring period. 
The variation is of the order of $\sim$11\% between the
SDSS spectra obtained during epochs 1 and 2 ($\sim$800 days in the quasar rest frame) and $\sim$17\% between the SALT spectra of epochs 5 and 7 ($\sim$ 240 days in the quasar's rest frame). 

Reverberation mapping studies of high redshift, high luminosity quasars have yielded a BLR size - luminosity relationship using the time delay between the continuum and the \civ \ emission line flux variations \citep[refer to equation 2 of][]{kaspi2007}. Using this relationship and the quasar luminosity $L_{\lambda}(1350 \si{\angstrom})= 2.56\times 10^{43}$ erg s$^{-1}$ \si{\angstrom}$^{-1}$ measured from the epoch-1 SDSS spectrum, we estimate the \civ \ BLR size of \J16 \ to be $0.10 \pm 0.04$ pc. 
%
Thus we expect the response of the \civ\ emission line to the continuum variations to have a typical delay of up to 240 days in the quasar's rest frame (i.e $\sim$750 days in the observer's frame). It is also expected that emission line variations will be diluted due to the extended nature of the BLR.

The CRTS and r-band lightcurves discussed in Section~\ref{sec:lc} show that apart from an early long term fading there is no clear long-term variability (by more than 0.2 mag) in r-band magnitudes during our monitoring period. In particular over epochs 5, 6 and 7 spanning 746 days (i.e 240 days in the quasar frame), during which we observe \civ\ flux variations by 17\%, the r-band flux has not changed by more than 0.1 mag. 
{\it Thus it appears that the variations in the flux of ionizing photons is decoupled from that of the r-band flux. Also, the ionizing flux needs to decrease between epoch-5 and 7 in order for the \civ\ line flux to decrease. Alternatively the BLR in \J16 is dynamically evolving without reaching a proper steady state equilibrium as assumed in reverberation mapping studies. }

%
%

Although their numbers are increasing in the recent times, \civ\ reverberation mapping studies of high-$z$ high luminosity quasars are still rare. \citet{Lira2018} have monitored 17 targets and found that three of them show unexpected line variability (i.e no clear correlation between r-band magnitude and emission line variabilities and/or some emission lines do not respond to continuum variations at all).
In such cases it is clear that the 
the r-band continuum variability do not mirror the ionizing continuum variations.
Even in the well studied NGC~5548 there are periods over which abnormal variability is noticed \citep{Goad2016}. The latter authors suggest that this abnormal variability is either due to changes in the intrinsic SED or is the consequence of high energy photons obscuration by the intervening gas. This may be what happens in the case of \J16.





\subsection{Absorption line variability:}
\label{sec:PI}

The BAL variability is usually understood in terms of ionization changes and/or bulk motions perpendicular to our line of sight.
The emergence of the blue-BAL and the emergence and subsequent disappearance of component-C of the red-BAL emphasize the importance of bulk motions in the case of \J16. 
On the other hand, there are indications that the ionizing radiation may be changing based on the \civ\ emission line flux variations. Thus it is most likely that the 
overall correlated variability we notice in section~\ref{sec:abs} could be driven by both these effects. While capturing a dynamical outflow by keeping track of appropriate radiative transport is difficult \citep[see for example,][]{Higginbottom2014} and  beyond the scope of this work, in the following we investigate simple photoionization models 
and 
1D disk-wind models in order to interpret our results.

First, we construct photo-ionization models using {\sc Cloudy} v17.01 \citep[][]{cloudy2017} to quantify the ionization induced variations.
Well constrained photoionization modelling is possible only 
when several species can be discussed. The detection of Si~{\sc iv} and Al~{\sc iii} absorption in the blue-BAL and Si~{\sc iv} during epoch-4 for the red-BAL provide this opportunity.
 However, due to the low resolution spectra used here, it will be difficult to measure the column densities accurately as hidden line saturation and unknown covering factor can lead to underestimate the column densities. It is possible that the covering factors are different for different ions that will also make the observed equivalent width ratios not following the intrinsic column density ratios. Also a simple geometry considered in {\sc Cloudy} may not be a correct representation of a multi-phase outflow.
 Given all these caveats, we use the model results as a rough guideline to infer the nature of the absorption line variability.

\subsubsection{Basic Cloudy model:}
\begin{figure*}
\centering
\includegraphics[viewport=65 5 1270 610,width=0.8\textwidth,clip=true]{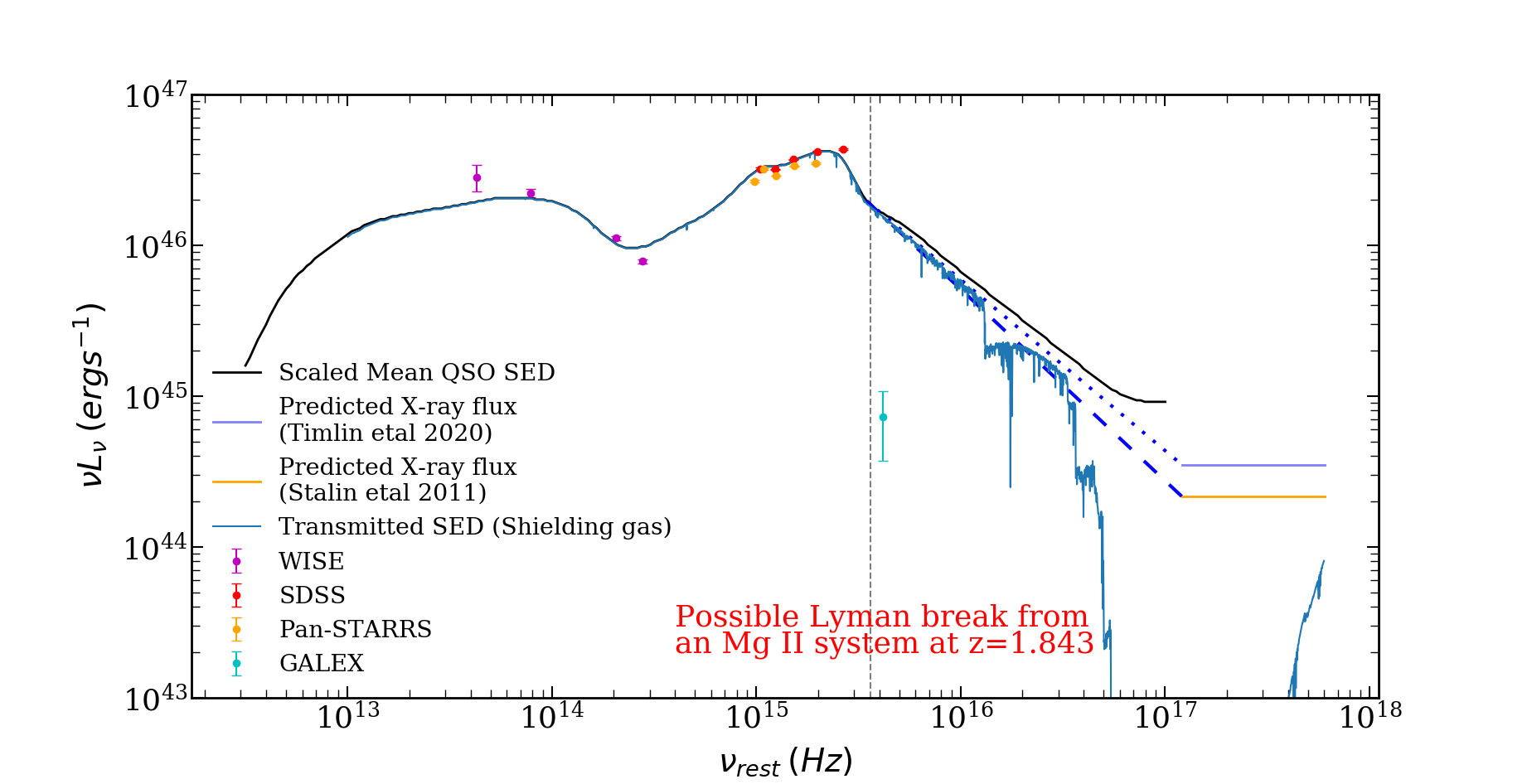}
\caption{Broad band spectral energy distribution of \J16 created using the mean quasar SED from \citet{richards2006} and photometric points taken from SDSS, WISE and GALEX. Pan-STARRS measurements are also shown.
They are slightly lower than the SDSS measurements (see Section~\ref{sec:lc} for variability). We extended this SED to the high energies using the observed relationship between L$_{2500}$ and $\alpha_{OX}$ for non-BAL quasars \citep{timlin2020} and BAL quasars \citep{stalin2011} and these are shown with blue dotted and dashed lines respectively. The figure also shows the non-BAL SED (i.e  blue solid line) after transmission through a shielding gaseous slab (see Section~\ref{sec:PI} for details). The GALEX measurement (cyan point) is well below our predicted SEDs. We believe this is because of Lyman continuum absorption produced by the $z = 1.843$ intervening \mgii\ absorber. The position of the Lyman limit of this absorber in the quasar rest frame is shown by the dashed vertical line. }

\label{sed}
\end{figure*}

We approximate the absorbing gas to be a plane parallel slab of uniform density, $\mathrm{n_H}$, 
having solar metallicity and illuminated from one side by the quasar radiation. The gas temperature and ionization states are computed in {\sc Cloudy} under equilibrium conditions. We consider the gas to be optically thin to H~{\sc i} ionization \citep[see for example,][]{Hamann1997}.

First we construct the observed wide band spectral energy distribution (SED) of \J16 using photometric points from the Wide-field Infrared Survey Explorer (WISE), SDSS and Galaxy Evolution Explorer (GALEX) in the infrared, optical and UV ranges, respectively (see Fig.~\ref{sed}). 
%
It is clear from this figure that the observed luminosities in the rest frame UV-to-optical range is well reproduced by the mean non-BAL quasar SED of \citet{richards2006}. However, this predicts much higher flux in the NUV band of GALEX compared to what is observed (cyan point in Fig.~\ref{sed}). We believe this is mainly due to the 
H~{\sc i} Lyman limit of strong intervening Mg~{\sc ii} systems found at \zabs = 1.84182 and 1.84201 with associated 
absorptions from \mgi, C~{\sc ii}, C~{\sc iv}, Si~{\sc ii}, Si~{\sc iv}, Al~{\sc ii}, Al~{\sc iii} and Fe~{\sc ii} detected.
%
In addition we do see a strong \lya\ absorption (with W(\lya)$\sim$ 10\AA) at the expected position in our NTT spectrum. However, poor SNR in this wavelength range prevents us from measuring the H~{\sc i} column density accurately.
%

%
%
As X-ray measurements are not available for \J16,  we have constructed the UV to X-ray part of the SED using the best-fit relationship between $\alpha_{ox}$ and monochromatic luminosity at rest 2500 \AA ($L_{2500}$) as given in \citet{timlin2020} for a sample of 2106 radio-quiet quasars from the SDSS using Chandra observations. 
This gives the monochromatic luminosity at 2 KeV, ${\rm L_{2~KeV}}$. This together with the assumption that $L_\nu \propto \nu^{-1}$ in the X-ray range allows us to construct the optical to X-ray SED as shown in Fig.~\ref{sed} (solid blue line). 
%
It is well known that quasars with BAL outflows tend to have lower X-ray emission compared to non-BAL quasars of similar $L_{2500}$ \citep{gibson2009}. In order to accommodate this, a similar procedure was carried out using the relationship between optical luminosity and $\alpha_{ox}$ obtained from a sample of BAL quasars as given in \citet{stalin2011} (orange solid line in Fig.~\ref{sed}).
%

\begin{figure}
    \centering
    \includegraphics[viewport=35 15 1000 650,width=0.47\textwidth,clip=true]{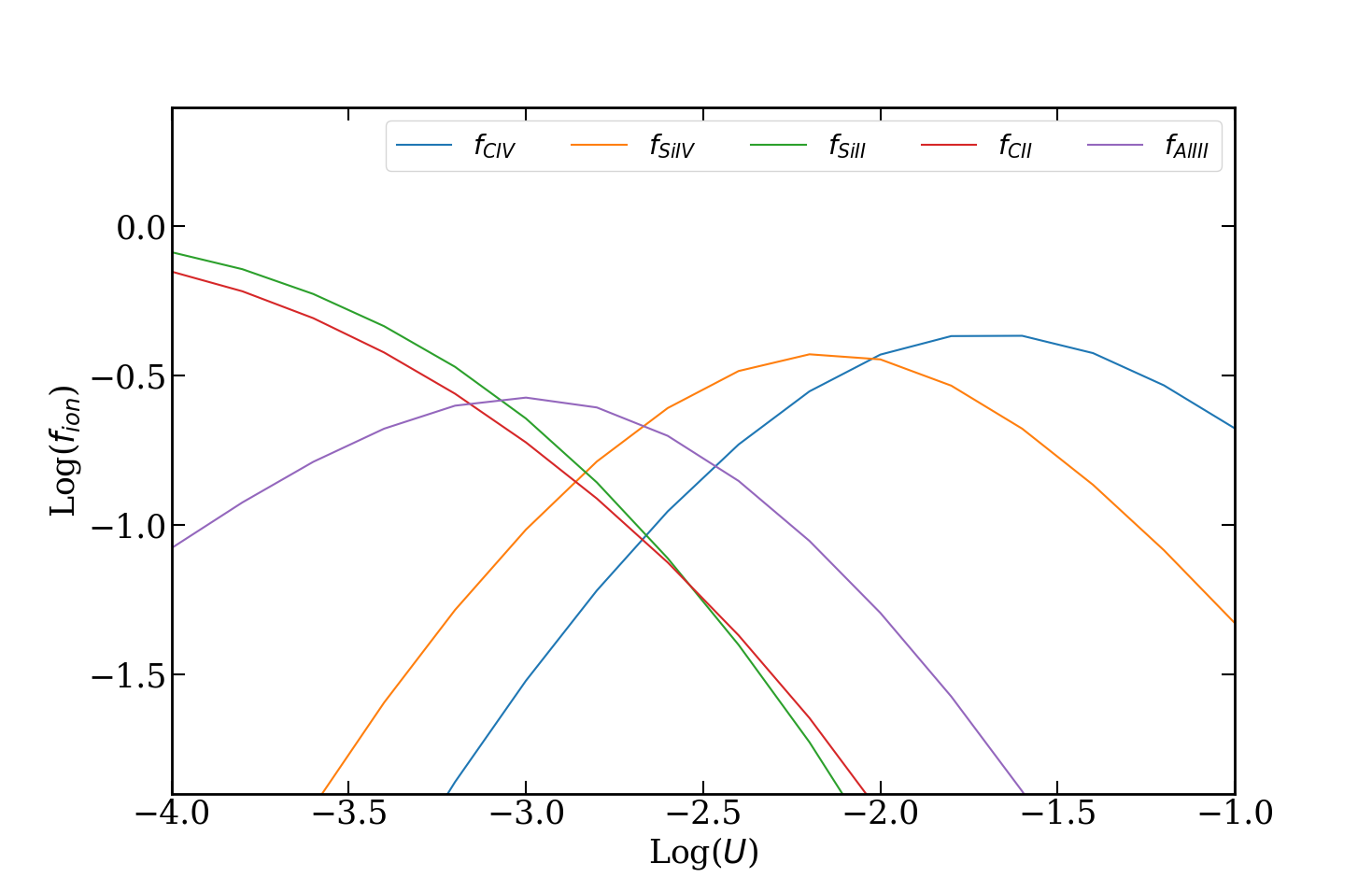}
    \caption{Ion fraction ($f_{ion}$) predicted by our models as a function of ionization parameter ($U$). The absorbing gas is assumed to be optically thin to H~{\sc i} ionizing photons. We have used the "transmitted SED"
    shown in Fig.~\ref{sed} as the ionizing radiation in our models.
    }
    \label{reda_cloudy}
\end{figure}


We consider a non-BAL SED modified by the presence of shielding gas or failed wind as suggested by some of the wind models \citep{murray1995, proga2000} 
as the ionizing radiation (also shown in Fig.~\ref{sed}). 
%
We run {\sc Cloudy} twice to produce this SED. In the first {\sc Cloudy} run, a continuum with the modelled non-BAL SED of \J16 is incident on the shielding gas with a high ionization parameter, log$(U)= 1$ (since the shielding gas is located close to the source), total hydrogen column density $N(H) = 10^{23}$ cm$^{-2}$ and solar metallicity. We obtain the transmitted continuum after it passes through the shielding gas. 
In the subsequent {\sc Cloudy} runs, this transmitted continuum is used as the ionizing radiation for the absorbing gas. For the assumed SED the number of ionizing photons per second from \J16 is, Q $\sim 3.4\times 10^{56}$ s$^{-1}$. In this case the distance of the absorbing gas ($r$) and ionization parameter (U) are related by,
\begin{equation}
{r \rm = 154.6 \times \bigg{(}{10^5\over n_H}\bigg{)}^{0.5}~~ \bigg{(}{10^{-1.4}\over U}\bigg{)}^{0.5} }~~~{\rm pc}.
\label{eqn:u}
\end{equation}
%
%
%
%
%
In  Fig.~\ref{reda_cloudy}, we show the ion fractions as a function of ionization parameter. 

\subsubsection{Blue-BAL absorption: ionization}
Emergence of the blue-BAL was detected during epoch-2. In addition
the \civ\ equivalent width has initially increased 
and subsequently decreased
after epoch-4 (see Table~\ref{tab_gauss} and Fig.~\ref{fig:absvar}). If these changes are purely driven by ionization changes we would like to have log~U $\sim$-1.5 for epoch-4 (i.e for which the \civ\ ion fraction peaks).
If we assume all the ions have similar covering factor and the gas composition is of solar abundance then we will be able to
infer whether log~U is more or less than $-1.5$ during other epochs.
In the epoch-2 spectrum we detect \civ, \siv\ and Al~{\sc iii} absorption (see Fig.~\ref{fig:bluesi4}). The fact that Al~{\sc iii} equivalent width is a factor of 2 less than that of \siv\ is consistent with log~U$>-$2.7 (see Fig.~\ref{reda_cloudy}).
 Here we have assumed the lines are not saturated and used the linear part of the curve of growth to estimate the column density ratios from the equivalent width ratios. Also relative abundances of metals are assumed to be that of solar values. The observed Al~{\sc iii} optical depth is lower than that of Si~{\sc iv}. Therefore, in the scenario where there is partial coverage we expect the effect of saturation to be much higher for Si~{\sc iv}, that is why we consider the above mentioned log~U to be a lower limit.
Similarly, to have \civ\ equivalent width twice that of \siv\ we need log~U$\ge-$2.2.  We consider it as a lower limit as the \civ\ profile is much deeper than the \siv\ profile.

Between epoch-2 and  3 the \civ\ equivalent width remained nearly the same while the \siv\ equivalent width has reduced by a factor 2. This indicates an increase in the ionization parameter between epoch-2 and 3.
Compared to epoch-2, the \civ\ equivalent width has increased by a factor of 1.5 in the spectrum taken during epoch-4. The Al~{\sc iii} absorption is not detected in the epoch-4 spectrum with a 3$\sigma$ upper limit of 0.32\AA. This trend is possible if the ionization parameter is increased between epoch-2 and 4. If the changes in $\mathrm{n_H}$ and $r$ are negligible between these epochs (however see below) then increase in ionization parameter will also imply increase in the ionizing flux.
Such an evolution in the ionization flux will also explain the increase in the \civ\ broad emission line equivalent width between these epochs.
As we have seen before the \civ\ equivalent width starts to decrease after epoch-4. This will imply reduction in the ionization parameter. Unfortunately our spectra do not cover Si~{\sc iv} or Al~{\sc iii} during the last 3 SALT epochs to confirm whether we detect variation in the opposite direction for Si~{\sc iv} and Al~{\sc iii} absorptions. However, this will be consistent with what is required to understand the \civ\ emission line flux variations.

It thus seems that we have a broad coherent picture.
However if variations in the ionizing radiation is the only driver for the equivalent width variations then we would have expected the restoration of the absorption line profiles for a given equivalent width. This is not the case as can be seen from Fig.~\ref{fig:absvar}. Thus one has to invoke also possible kinematic shifts (i.e acceleration) in the absorbing gas.

\subsubsection{Blue-BAL absorption: kinematics}

Since the blue-BAL component has emerged only during epoch-2 and shows signatures of acceleration till epoch-4 we can assume it to be located close to the accretion disk. 
If true, using the velocity and acceleration as a function of time we can place constraints on the exact location of the absorbing gas.

Radiatively accelerated wind models predict a radial velocity profile of the form \citep[see][]{murray1995},
\begin{equation}
    v(r) = v_{\infty} \Big(1-\frac{r_f}{r} \Big)^\beta 
\end{equation}
where $v_{\infty}$ is the terminal velocity, $\beta \sim 1.15$ and $r_f$ is the launching radius of the wind. Then, the acceleration profile is given by,
\begin{equation}
    a(r) = v \frac{dv}{dr} = \beta v_{\infty}^2 \frac{r_f}{r^2} \bigg( 1 - \frac{r_f}{r} \bigg) ^{2\beta - 1}.
\end{equation}
Once we fix $r_f$ and $v_\infty$ we can predict $v(t)$ and $a(t)$ for a given fluid element.
We can fit the velocity and centroid shift measured at different epochs using the above two equations.
\begin{figure}
\includegraphics[viewport=30 20 1200 650,scale=0.23,clip=true]{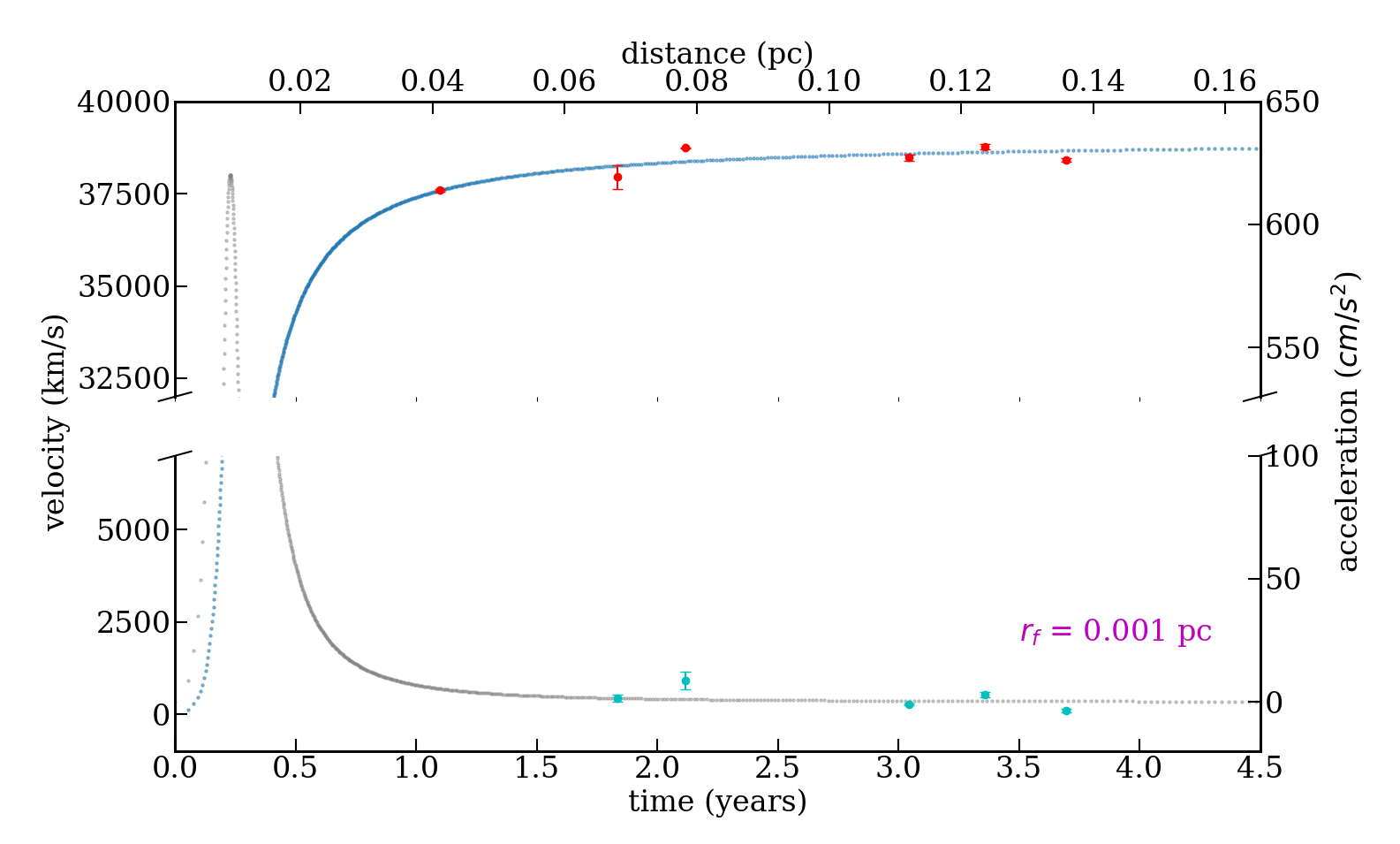}
\caption{The velocity (blue dotted lines) and acceleration (gray dotted lines) profiles as a function of $r$ and $t$ are shown assuming $v_{\infty} = 39000$ kms$^{-1}$ and $\beta \sim 1.15$. }
\label{vel_law}
\end{figure}

First we consider that the absorbing gas was ejected from the place where the Keperian velocity is close to the terminal velocity. 
For \J16, we obtain $M_{\rm BH}$ to be $(3.66 \pm 0.15) \times 10^9 M_\odot$ using the FWHM 
of the \civ\ emission (FWHM = 5977 $\pm$ 123 kms$^{-1}$), the monochromatic continuum luminosity, $L_{\lambda}(1350 \si{\angstrom})= 2.56\times 10^{43}$ erg s$^{-1}$ \si{\angstrom}$^{-1}$, measured from the first epoch SDSS spectrum (i.e MJD of 53501) and the empirical mass-scaling relationship given by \citet{vestergaard}.
%
%
We obtain $r \sim 0.4$ pc and $t \sim 11$ years since the launching of the flow to reach the observed line of sight velocity in epoch-2. 
Using log~U $\sim$-1.4 and r = 0.4~pc we get $\mathrm{n_H \sim 1.5\times10^{10}~cm^{-3}}$ using Eqn.~\ref{eqn:u}. However in this case
we do not expect acceleration of the gas beyond epoch-2. This is inconsistent with our data. In order to accelerate the gas at latter epoch one has to perturb the steady state solution assumed here.

%
%


We therefore carefully choose the launching radius of the flow 
in such a way that $v_{obs}$ and $a_{obs}$ occurs at the same distance (r) (and hence at the same time also).
In this case at $r \sim 0.040$ pc and $t \sim 1$ year at epoch-2, this model gives epoch-2 $v_{obs}$ = 37591 kms$^{-1}$ and acceleration values of a few cms$^{-2}$ which can roughly reproduce the observed changes in velocity in the subsequent epochs satisfying the observed time-scales (see Fig.~\ref{vel_law}). Also in this scenario the emergence occurs much after the first spectroscopic epoch. The consequence of the smaller launching radius are: (1) the required gas density constrained by the ionization parameter will be, $\mathrm{n_H \sim~ few~\times10^{12}}$ cm$^{-3}$; (2) the displacement of the gas between epoch-2 and 4 is $\sim$0.04~pc. 
As this is a factor 2 change in distance from the quasar one has to take into account the change in ionization parameter from the gas motion also.

Although appealing, as pointed out by \citet{grier2016}, such models can reproduce the observed acceleration between two epochs but they cannot reproduce small jerks. While the overall velocity and acceleration evolution is capture in Fig~\ref{vel_law} small changes as seen between epoch 5 to 7 are not reproduced. They could be due to instabilities in the flow.

\subsubsection{Red-BAL absorption:}

As we have seen before the largest \civ\ equivalent width for both A and B components were observed during epoch-4. As we do not detect any other species from component-B it is difficult to constrain model parameters. So we focus mainly on component-A.  The \siv\ ion fraction is maximum at log~U$\sim-2.2$. If we  assume solar abundance ratio for the metallicities, we need log~U$\ge-2.6$ in order to produce the observed \civ\ to \siv\ equivalent width ratio. The limit is 
obtained as explained in the previous section.
%
In this case a slow increase in the \civ\ equivalent width with a large increase in the \siv\ equivalent width would be consistent with ionization parameter, log~U$\ge-1.5$, with a decrease between epoch-2 and 4. As both red- and blue-BAL components see the same radiation field the opposite requirements for the ionization parameter is inconsistent with the simple picture of no change in $\mathrm{n_H}$, $r$ and probably total amount of gas along our line of sight (i.e $N$(H)).

If the observed ejection velocity is related to the terminal velocity then red-BAL component will be ejected at a larger $r_f$ (i.e $\sim$6 times more) compared to the blue-BAL. Recall, components A and B emerged during epoch-2 from a broad single component seen during epoch-1. From the bottom plots of Fig.~\ref{fig:wvar}, we see the centroid velocity of A and B components remain nearly constant between epochs 2 and 4. There is signature of acceleration between epochs 4 and 5 before staying at constant velocity. Such a change is velocity as cannot be naturally explained by the time-steady disk wind model discussed above. One has to invoke instabilities to trigger such acceleration after a nearly constant velocity phase.


\subsubsection{Possible scenario of correlated variability:}

The large outflow velocities we see for the \civ\ gas can originate from disk-winds driven by either radiation pressure \citep[for example,][]{arav1994} or magnetic driving \citep[see for example,][]{dekool1995}.
In the line-driven wind models the low ionization gas (shielded from direct heating from the central X-ray source) can be accelerated to large velocities close to the equatorial plane. Thus the wind has a funnel shaped structure reaching a nearly steady state with  stable density and velocity profiles \citep{proga2000,Proga2004}. 
 While the disk-wind models are yet to mature to the level of explaining various observed variability signatures,
 \citet{Dyda2018} have shown that a time-varying radiation source can induce density and velocity perturbations in the acceleration zones of line-driven winds. It is also understood that larger
 the velocity of the gas, closer the gas will be to the central black hole.
While it is not clear what would be the inclination of the disk-wind in \J16\ relative to our line of sight,
it is reasonable to assume that the correlated equivalent width and kinematic changes could be triggered by fluctuations in the radiation field. In particular, RWB colour variations could suggest possible propagation of fluctuations from the outer to the inner part of the disk. Measuring the column densities of different species using high resolution spectroscopy would be important to constrain the locations of red- and blue-BAL components. This is an important step to make progress in the right direction.
 

 In the literature various possible mechanisms are proposed to understand the kinematic shift in the absorption lines. These include, directional shift of the outflow, rotation of the absorbing gas causing the line of sight velocity to change,  gravitational redshift, gas infall, binary quasars, interaction with an ambient medium etc.,
\citep{Hall2002, Gabel2003, Hall2013, grier2016,Joshi2019}.
Some of these (like interacting with ambient medium) work well to explain the deceleration seen in some BAL quasars. Mechanims such as gas infall, binary quasars and gravitational redshifts were introduced to understand broad absorption seen with redshifts greater than quasar redshifts.
Among the above listed  possibilities, the large outflow velocities and acceleration signatures seen in \J16\ can be explained with directional changes. Directional change can naturally occur in disk wind models where the gas trajectories are curved with respect to our line of sight probably due to magnetic field lines. With appropriate configuration it will be also possible to explain the correlated variability as well. Such a scenario needs to be explored in simulations.
On the other hand circular trajectories can also occur if the winds are ejected from rotating disks. However a simple rotation scenario may have issues to explain the large velocities and acceleration seen in the case of  the blue-BAL and the correlated variability seen between blue and red-BAL discussed here.


\section{Summary}
\label{sec:summary}

We have presented a detailed variability analysis of BAL quasar J162122.54+075808.4 (\zem = 2.1395) spectra obtained during 7 epochs spanning almost 15 years in the observer's time scale. This has allowed us to probe the absorption  and emission line variability over time-scales of $\sim$3 months to $\sim$4.8 yrs in the quasar's rest frame.

\J16 was identified in the SDSS survey as a BAL-QSO with \civ\ absorption at an ejection velocity of  
$v_{\rm e} \sim$ 15,400 km~s$^{-1}$, referred to as red-BAL in this work. We report the emergence of 
%
a new BAL component with associated absorption from \civ, \siv\ and Al~{\sc iii}) with $v_{\rm e} \sim$ 37,500 km~s$^{-1}$ in the second epoch spectrum, referred to as blue-BAL. 
We study the variability of these two BAL components using additional spectra we have obtained with NTT and SALT telescopes.

The rest equivalent width of the \civ\ absorption line of the blue-BAL shows an increasing trend followed by a steady decline. During the early phases the absorption profile shows kinematic shifts consistent with acceleration. The acceleration we derive between epoch-2 and 3 is $+8.84\pm3.61$ cm s$^{-2}$ and  epoch-2 and 4 is $+3.57\pm0.08$ cm s$^{-2}$. 
These are among the largest acceleration measured in BAL quasars \citep[see figure 6 of,][]{grier2016}. 
In the later epochs we do see some non-monotonic centroid shifts. Such trends are not predicted by the time-steady disk wind models but often seen in 
BAL quasars with acceleration signatures monitored for more than two epochs.


During our monitoring period the red-BAL shows substantial equivalent width and kinematic changes as well. The total \civ\ equivalent width increases in the initial stages (till epoch 4) then shows a decreasing trend consistent with what we saw for the blue-BAL component (see Fig.~\ref{fig:wvar}).
In particular the \civ\ absorption profile showed substantial variation.
In the beginning,
the \civ\ absorption profile is consistent with a single Gaussian component (i.e in the epoch-1 spectrum).
During our monitoring period the \civ\ absorption is seen in three distinct components (called A, B and C). Component C emerged during epoch-3 and disappeared in the following epochs. The time-scale for emergence and subsequent disappearence for this  component (i.e $\sim$269~rest frame days in quasar's frame) is one of the shortest measured among BAL quasars.
The equivalent width variations of individual components A and B follow that of the blue-BAL component. 
These components also show kinematic shifts consistent with acceleration between epochs 5 and 6. As in the blue-BAL, the velocity and acceleration as a function of time are inconsistent with a time-steady disk-wind model predictions.

The most interesting part of the absorption line variation is the presence of correlated variability between different absorption components that are well separated in velocity space. Usually such correlated variability is associated to photo-ionization induced 
variations. To explore this we investigate the time variability of the emission line fluxes using our spectra and of the continuum light using available photometric light curves.


%

The \civ\ emission line profile of \J16\ is asymmetric and consistent with the presence of outflows. We observe a clear long term time variation of the \civ\ emission line equivalent width.
We find that the \civ\ line flux varies by up to 17\% over the
time of the observations. 
In the standard stratified BLR picture, this would request even larger variations in the ionizing flux.
%
However, available optical (rest frame UV) light curves do not support any systematic brightening or fading of \J16 on the corresponding time-scales. The structure function constructed from the r-band light curve is consistent with \J16\ being less variable compare to normal SDSS quasars. 
Nonetheless, we do see g-r color variations with a ``red-when-bright" trend, the amplitude of the flux variations being larger at smaller wavelengths.
Thus if emission line variations are triggered by large ionization changes then they must be decoupled from the changes in the r-band flux. Alternatively 
\J16 is dynamically evolving without reaching a proper steady-state equilibrium as assumed in reverberation mapping studies.


We consider simple photoionization models to gain some insights into the density of the gas and its location.
Simple models with constant density and same location predict
the restoration of the 
\civ\ profile for a given 
equivalent width. Since this is not what we observe, 
we rule out simple ionization changes. Disk wind models suggest that the gas density of the blue-BAL has to be larger than 10$^{10}$ cm$^{-3}$ to produce consistent ionization together with velocity profiles and acceleration. The observation of non-monotonic changes in the acceleration and the lack of restoration of the absorption profile can be reconciled by the presence of density and velocity fluctuations in disk winds. Simulations do suggest that a time-varying radiation source can produce such perturbations.

Therefore, in the frame work of available disk-wind models we favor density and velocity field fluctuations triggered perhaps by varying radiation field (and associated disk-instabilities) to cause the observed variability in \J16.
Important progress could be made if we could constrain the density (and hence the location with respect to the central UV source) and the size of the absorbing gas components. 
For this, high resolution spectroscopic observations of this interesting source are needed in order to derive
accurate column densities and gas covering factors. 

    


\section*{Acknowledgements}
We thank  Nishant Singh,  Aseem Paranjape and K. Subramanian for useful discussions. PA thanks Labanya K Guha for helpful discussions on several python programming techniques used in this paper.
PPJ thanks Camille No\^us (Laboratoire Cogitamus) for 
inappreciable and often unnoticed discussions, advice and support.

\section*{Data Availability}
Data used in this work are obtained using SALT. Raw data will become available for public use 1.5 years after the observing date at https://ssda.saao.ac.za/.



\bibliographystyle{mnras}
\bibliography{mybib_bal} 



\bsp	
\label{lastpage}
\end{document}